\documentclass[twocolumn]{aa}  
\usepackage{graphicx,natbib,journal_names}
\usepackage[breaklinks,colorlinks,urlcolor=blue,citecolor=blue,linkcolor=blue]{hyperref}
\usepackage{multirow,afterpage,pdflscape,lipsum,capt-of,mathtools}

\usepackage[dvipsnames]{xcolor}
\usepackage{comment}
\usepackage{float}
\usepackage{txfonts}
\usepackage{soul}
\usepackage{array}
\usepackage{colortbl}
\definecolor{Grey}{rgb}{0.5,0.5,0.5} 
\definecolor{MyRed}{rgb}{0.9,0.0,0.0} 
\definecolor{MyPink}{rgb}{0.8,0.3,0.5} 
\definecolor{MyMediumBlue}{rgb}{0.7,0.72,1.0} 
\definecolor{MyGreen}{rgb}{0.0,0.9,0.5} 
\definecolor{SWGreen}{rgb}{0.0,0.5,0.0} 
\definecolor{SWRed}{rgb}{0.9,0.0,0.0} 
\definecolor{AMpurple}{rgb}{0.9,0,1.0} 
\definecolor{AMbrown}{rgb}{0.8,0.3,0.3} 

\usepackage{ulem}

\begin{document}

\title{Comparison of chromospheric diagnostics in a 3D model\\
atmosphere: H$\alpha$ linewidth and mm continua}
\titlerunning{Comparison between H$\alpha$ and mm continua}
\author{Sneha Pandit\inst{1,2}, Sven Wedemeyer\inst{1,2}, Mats Carlsson\inst{1,2}, Miko\l{}aj Szydlarski\inst{1,2}}
\institute{Rosseland Centre for Solar Physics, University of Oslo, Postboks 1029 Blindern, N-0315 Oslo, Norway
\and
Institute of Theoretical Astrophysics, University of Oslo, Postboks 1029 Blindern, N-0315 Oslo, Norway\\
\email{sneha.pandit@astro.uio.no}}

\abstract
  {
    The H$\alpha$ line, one of the most studied chromospheric diagnostics, is a tracer for magnetic field structures, while its line core intensity provides an estimate of the mass density. The interpretation of H$\alpha$ observations is complicated by deviations from Local Thermodynamic Equilibrium (LTE) or instantaneous statistical equilibrium conditions. Meanwhile, millimetre continuum radiation is formed in LTE, hence the brightness temperatures from Atacama Large Millimetre-submillimetre Array (ALMA) observations provide a complementary view of the activity and the thermal structure of stellar atmospheres. These two diagnostics together can provide insights into the physical properties like temperature stratification, magnetic structures, and mass density distribution in stellar atmospheres.
}
    {In this paper, we present a comparative study between synthetic continuum brightness temperature maps for millimetre wavelengths (0.3~mm to 8.5~mm) and the width of the H$\alpha$ 6565~\AA~line.}
  {
   The 3D radiative transfer codes Multi3D and Advanced Radiative Transfer (ART) are used to calculate synthetic spectra for the H$\alpha$ line and the mm continua respectively, from an enhanced network atmosphere model with non-equilibrium hydrogen ionisation generated with the state-of-the-art 3D radiation magnetohydrodynamics (rMHD)  code Bifrost. We use Gaussian Point Spread Function (PSF) for simulating the effect of ALMA's limited spatial resolution and calculate the H$\alpha$ vs. mm continuum correlations and slopes of scatter plots for the original and degraded resolution of the whole box, quiet sun and enhanced network patches separately.
}
  {
The H$\alpha$ linewidth and mm brightness temperatures are highly correlated and the correlation is highest at a wavelength 0.8~mm i.e. in ALMA Band~7. The correlation systematically increases with decreased resolution. 
On the other hand, the slopes decrease with increasing wavelength. The degradation of resolution does not have a significant impact on the calculated slopes.
 }
   {With decreasing spatial resolution the standard deviations of the observables, H$\alpha$ linewidth and brightness temperatures decrease and the correlations between them increase, but the slopes do not change significantly. Hence, these relations may prove useful to calibrate the mm continuum maps observed with ALMA.}

\maketitle

\section{Introduction}

One of the most commonly used chromospheric diagnostics is the H$\alpha$ line, i.e. the transition between atomic levels 3 and 2 in the hydrogen atom \citep{1985ApJ...294..626C}. Due to its high opacity in the line core, the H$\alpha$ line core forms in the low plasma beta regime, i.e. with more dominant magnetic pressure than the plasma pressure, which results in magnetic fields being the main structuring agent in the upper chromosphere \citep{2012ApJ...749..136L}. Hence, the H$\alpha$ line is a very good diagnostic for the magnetic structures in the chromosphere. The chromospheric mass density can be traced with the H$\alpha$ line core intensity \citep{2012ApJ...749..136L}. The low atomic mass of hydrogen results in high-temperature sensitivity of the H$\alpha$ line width through significant thermal Doppler broadening \citep{2009A&A...503..577C}. Even though the H$\alpha$ line is a useful diagnostic of stellar chromospheres, since it forms in non-local thermodynamic equilibrium (NLTE), it is not particularly suitable for the determination of the chromospheric temperature stratification \citep{1991A&A...251..199P}.

Based on radiative transfer calculations for a  solar atmospheric radiation-magnetohydrodynamics simulation \citep{2016A&A...585A...4C}, \citet{2012ApJ...749..136L} show that the H$\alpha$ line width is correlated with the temperature of the plasma while the line core intensity is a diagnostic of the mass density in the chromosphere. These relations come from the fact that the H$\alpha$ line is a strongly scattering line such that the source function is non-locally determined by the mean intensity. A high local temperature means that the atomic absorption profile gets wide (due to the low atomic mass, the width of the profile is dominated by thermal broadening) and since the source function is rather insensitive to the local conditions, we also get a wide intensity profile. A high mass density in the chromosphere leads to a line core formation at a larger height where the source function is lower, hence a lower line core intensity.

\begin{figure*}[!tp]
\centering
\includegraphics[width=1.0\textwidth]{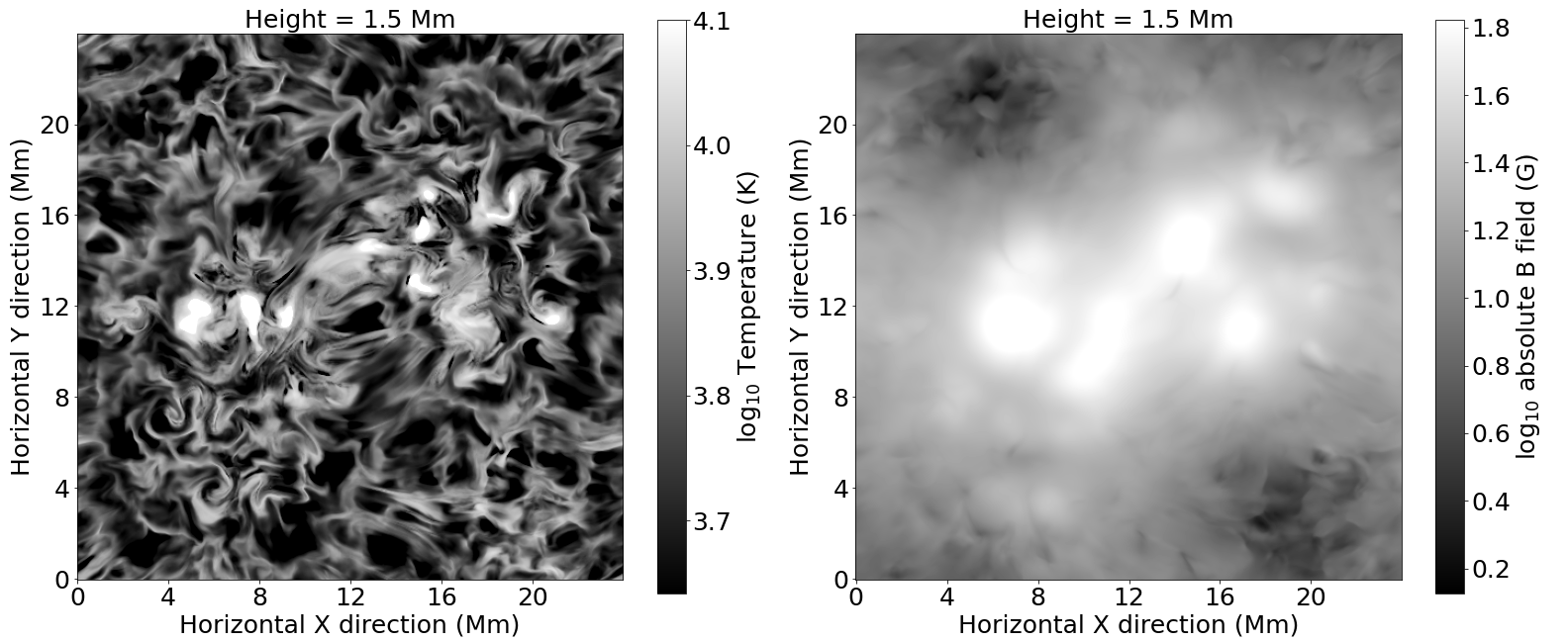}
\vspace{-4mm}
\caption{Horizontal cross-section of temperature (left) and the absolute magnetic field ($|B|=\sqrt{B_x^2+B_y^2+B_z^2}$) (right) from a snapshot of the Bifrost simulation at a constant height of 1.5\,Mm from the surface.}
\label{fig:Bifrost_B_T}
\end{figure*}

It is well established that the continuum radiation at millimetre (mm) wavelengths is formed at chromospheric heights \citep[see, e.g.,][and references therein]{2016SSRv..200....1W}. The continuum radiation at millimetre wavelengths (here 0.3–8.5~mm) originates from free-free emission in the chromosphere, and the two main opacity sources are  
H and H$^-$ free-free absorption
\citep{1985ARA&A..23..169D}. These processes result in a Local Thermodynamic Equilibrium (LTE) source function as they are coupled to the local property of the plasma: the electron temperature. Hence, we can use the Rayleigh-Jeans law and interpret the emergent intensity (the observed brightness temperature ($T_\mathrm{b}$)) in the millimetre wavelength domain as local electron temperature \citep{2015A&A...575A..15L}. Thus, the millimetre continuum is a very convenient diagnostic to understand the temperature stratification in solar and stellar atmospheres \citep{2007IAUS..239...52W, 2020A&A...635A..71W}.

The H$\alpha$ line opacity is proportional to the hydrogen $n = 2$ level population; the H$\alpha$ optical depth scales with the $n = 2$ column density \citep{2007A&A...473..625L}. The availability of free electrons is dependent on the n=2 population of hydrogen in the upper atmosphere \citep{2019ApJ...881...99M}. The formation of the H$\alpha$ and mm continuum is influenced by the plasma temperature, the H(n=2) level populations and electron populations.

Traditional H$\alpha$ activity indicators, like the linewidth, Full Width at Half Maximum (FWHM) or integrated fluxes depend on several wavelengths across the line profile \citep{1989Ap&SS.161...61H, 2002AJ....123.3356G, 2014MNRAS.444.3517M, 2019ApJ...881...99M}. The H$\alpha$ line core is formed higher in the atmosphere than the wings. 
Therefore, while calculating these activity indicators from observations, we inherently look at a large volume of plasma in the chromosphere \citep{2009A&A...499..301V}, which makes 
determining the thermal stratification difficult. 
In this regard, the millimetre continuum has an additional advantage as a complementary diagnostic, namely that ---  to a first approximation --- the radiation at a given millimetre wavelength originates from more or less the same height range with the average formation height range increasing as a function of wavelength. 
While this assumption is probably valid on average, we note that the exact formation height ranges might show variations that are currently investigated 
 \citep[see, e.g.,][and references therein]{2021A&A...656A..68E,wedemeyerfrontiers, 2022ApJ...933..244H}.
On the other hand, the formation height of the H$\alpha$ line is in the range of 0.5~Mm to 3~Mm above the photosphere, which is a large range \citep{2012ApJ...749..136L}. 

The Atacama Millimetre/submillimetre Array (ALMA) \citep{2009IEEEP..97.1463W} provides observations with a high spatial and temporal resolution for the mm continuum. 
The H$\alpha$ linewidth and mm continua both depend on the temperature and the electron density of the volume of plasma where they are formed. 
This work is focused on using these two as complementary diagnostics similar to the case study on co-observational data for ALMA Band 3 and IBIS in \citet{2019ApJ...881...99M}. 
In this paper, we compare the mm continuum and H$\alpha$ linewidth synthesised from a 3D realistic Bifrost model atmosphere to understand how these can be used as complementary diagnostics for the chromosphere. The model and the spectral synthesis are described in Sect.~\ref{Sec:Methods}. In Sect.~\ref{Sec:Results}, we present the quantitative comparison of the two diagnostics which is further qualitatively discussed in Sect.~\ref{Sec:Discussion}. The study is concluded in Sect.~\ref{Sec:Conclusion}.

\section{Methods}
\label{Sec:Methods}

A snapshot of a numerical 3D simulation of the solar atmosphere (see Sect.~\ref{sec:Bifrost}) is used as input for radiative transfer codes to compute the H$\alpha$ line intensity (see Sect.~\ref{Multi3D}) and the continuum intensity at millimetre wavelengths (see Sect.~\ref{sec:ART}). 
The resulting maps for different wavelengths are then degraded to a spatial resolution 
typically achieved with ALMA as described in Sect.~\ref{sec:psf} before the H$\alpha$ line width is determined and compared to the mm brightness temperatures. 

\subsection{3D model atmosphere}
\label{sec:Bifrost} 

Bifrost is a 3D radiation magnetohydrodynamics (rMHD) simulation code which includes physics relevant to chromospheric conditions \citep{2011A&A...531A.154G}. 
The model used for this study is taken from a continuation of the enhanced network simulation en024048 at 17~min after the publicly released snapshots from \citet{2016A&A...585A...4C}.
The computational box of this model extends 24\,Mm $\times$ 24\,Mm horizontally, 
and vertically from 2.4\,Mm below the visible surface to 14.4\,Mm above it, thus encompassing the upper part of the convection zone, the photosphere, chromosphere, transition region and corona. The computational box consists of 504 $\times$ 504 $\times$ 496 grid points with a horizontal resolution of 48\,km, which corresponds to an angular size of approximately 0.066\,arcsec. The grid is non-equidistant in the vertical direction varying from 19\,km in the photosphere and chromosphere up to 5\,Mm height and then increasing to 100\,km at the top boundary. Both the top and bottom boundaries are transparent, whereas lateral boundary conditions are periodic. 
At the bottom boundary, the magnetic field is passively advected with no extra field fed into the computational domain. 
For further details see \citet[][and references therein]{2016A&A...585A...4C}. 

The temperature and absolute magnetic field showing the loops of the enhanced network with the quiet region at typical chromospheric heights are shown in Fig.~\ref{fig:Bifrost_B_T}. 
As seen in  Figs.~11 and 12 in \citet{2016A&A...585A...4C}, the simulation has different regions. The en024048 simulation features an enhanced network (EN) patch with loops in the middle of the computational domain as seen in Fig.~2 in \citet{2015A&A...575A..15L} or
(see the black square in Figs.~\ref{fig:H_a_ALMA_3mm_comparison_leenaarts_formula}~and~\ref{fig:H_a_ALMA_0.8mm_comparison_leenaarts_formula}), whereas the outer parts are more representative of magnetically less active,  quiet Sun (QS) conditions.  
In the absolute magnetic field slice through 1.74~Mm  (Fig.~\ref{fig:Bifrost_B_T}b), the footpoints of the magnetic loops are seen. The gas temperatures at those heights are comparatively lower at locations with higher magnetic strengths. And, the loops show significantly higher temperatures in the range of 2 to 3~MK from the footpoint to the apex of the loop.

\begin{figure}[tp!]
\vspace{-30mm}
\centering
\includegraphics[width=0.45\textwidth]{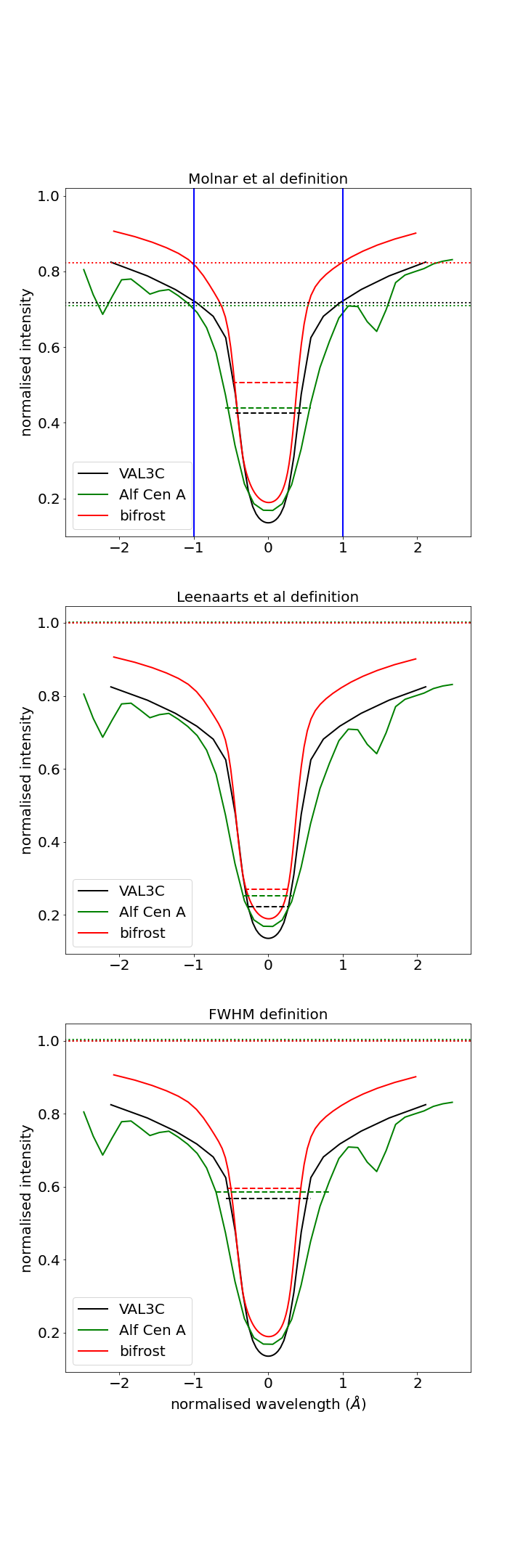}
\vspace{-25mm}
\caption{Comparison of the linewidth as defined by \citet{2019ApJ...881...99M}
(upper panel),  by \citet{2012ApJ...749..136L} (middle panel) and as defined as the FWHM (lower panel). The red, black and green lines show normalised spectra for the VAL~C model, Bifrost~QS and the observed spectrum for $\alpha$~Cen~A \citep{2008A&A...488..653P}, respectively. The solid blue lines in the top panel mark $\pm 1 \AA$, while the dotted lines show the considered maxima and the dashed lines show the calculated linewidths in all three cases.}
\label{fig:line_profiles_linewidth_definition_comparison}
\end{figure}

\subsection{H$\alpha$ spectral line synthesis} 
\label{Multi3D}

Multi3D \citep{2009ASPC..415...87L} employs the accelerated lambda iteration method developed by \citet{1992A&A...262..209R} with the extension to treat effects of partial frequency redistribution using the angle averaged approximation by \citet{2001ApJ...557..389U}. We used a five-level plus continuum hydrogen model atom. 
Microturbulence is not introduced, as \citeauthor{2012ApJ...749..136L} show that the synthesised spectra are similar to observational spectra even without using microturbulence. Further details of synthesising the H$\alpha$ spectra in 3D using Multi3D can be found in \citet{2012ApJ...749..136L}.

\subsection{H$\alpha$ observations of $\alpha$ Cen A}

In this study,  H~$\alpha$ observations of the solar-like star $\alpha$~Cen~A are used for comparison. The data is acquired from \citet{2008A&A...488..653P} and \citet{2005A&A...431..329L}. The H$\alpha$ observations are collected at Observatório do Pico dos Dias (OPD), operated by the Laboratório Nacional de Astrofísica (LNA), CNPq, Brazil on the ESPCOUDE 1.60m telescope. The data are wavelength-calibrated, Doppler-corrected, and flux-normalised to unity by \citeauthor{2008A&A...488..653P}.

\subsection{Definition of linewidth}
\label{sec:deflinewidth}

As this study is based on forward modelling, the linewidth definition by \citet{2012ApJ...749..136L} is used. However, we note that different definitions of the H$\alpha$ linewidth are used in the literature. The core width definition by \citet{2012ApJ...749..136L} and the definition used by \citet{2009A&A...503..577C} and  \citet{2019ApJ...881...99M} are compared to the Full Width Half Maximum (FWHM) definition for the linewidth in Fig.~\ref{fig:line_profiles_linewidth_definition_comparison}

The \citet{2012ApJ...749..136L} H$\alpha$~line-core width calculated for the Bifrost model as the full width at $1/10^{th}$ maximum, i.e. the separation between the line wings at $I=I_{min} + 0.1 (I_{max} -I_{min})$.
The representative wavelengths for the red and blue line wing that span the core width are calculated by first identifying the two spectral sampling points which are closest to the $1/10^{th}$ of the intensity between the pseudo continuum and the line core and then using linear interpolation.

\citet{2019ApJ...881...99M} follow the definition by \citet{2009A&A...503..577C} for their observational study. There, the linewidth is defined as the separation of the line profile wings at half of the line depth, where the maxima are defined at $\pm 1 \AA$ from the line core (referred to as  \citet{2019ApJ...881...99M} definition of linewidth, here onward). The line core intensity calculated for the Bifrost model used in this study reveals the structure of the EN loops. The variation in the activity in the model atmosphere among QS and EN regions is also visible from the variation of the H$\alpha$ line core intensity. It is to be noted that throughout the paper, the wavelengths mentioned are vacuum wavelengths.

The resulting line core width according to the \citet{2012ApJ...749..136L} definition is shown in Fig.~\ref{fig:H_a_ALMA_3mm_comparison_leenaarts_formula}a.
Please note that the plotted value ranges for these maps have been limited to the 99th to 1st percentile for better visibility, in particular, to highlight the imprints of the loop structures in contrast to the less active surroundings.

\subsection{Synthesis of millimetre continuum brightness temperatures} 
\label{sec:ART}

The intensity of the continuum radiation at millimetre wavelengths for the 3D input model (see Sect.~\ref{sec:Bifrost}) is calculated with the Advanced Radiative Transfer (ART) code by \citet{2021_art}. The output intensities $I_{\lambda}$ are then converted to brightness temperatures $T_b$ by using the Rayleigh-Jeans approximation: 
\begin{equation}
    T_b=\frac{\lambda^4}{2k_Bc}I_{\lambda},
\end{equation}
where, $\lambda, k_B$ and $c$ are the wavelength and Boltzmann constant and the speed of light,  respectively. 
In total, maps for 34 wavelengths are computed, covering the range that in principle can be observed with ALMA. That includes both currently available receiver bands and bands that might be offered for solar observations in the future. 
The resulting maps for the wavelengths of 3.0\,mm and 0.8\,mm are presented in Figs.~\ref{fig:H_a_ALMA_3mm_comparison_leenaarts_formula} and \ref{fig:H_a_ALMA_0.8mm_comparison_leenaarts_formula}, respectively, while further maps for selected wavelengths are shown in Sect.~\ref{Appendix_B}  (Figs.~\ref{fig:H_a_ALMA_0.4mm_comparison}-\ref{fig:H_a_ALMA_8.5mm_comparison}).

\subsection{Image degradation}
\label{sec:psf}

In principle,  the synthesised beam for interferometric observations with ALMA, i.e. the effective point spread function (PSF), depends on several factors such as the angle of the Sun in the sky with respect to the array baselines and the configuration of the antennas. The resulting beam would be typically elliptical, while the Fourier space of the source (here the Sun) would only be sampled sparsely 
\citep[see, e.g.,][]{1974A&AS...15..417H, 2013ExA....36...77B}.
For this theoretical study, circular Gaussians are used as PSFs as a first approximation as this simplified approach suffices to demonstrate the effect of spatial resolution. The more detailed modelling of the ALMA beam would depend on the exact array configuration, receiver band (some of which are not used for solar observations yet), and details of the imaging process, which would go beyond the scope of this study \citep{wedemeyerfrontiers}. 

As mentioned above, the exact shape and width of the synthetic beam varies even for observations in the same receiver band. For instance, the data sets found on the Solar ALMA Archive \citep[SALSA,][]{2022A&A...659A..31H} have widths around 2'' including the beam with a width of 1.92''\,$\times$\,2.30'' for the data set 2016.1.01129.S. That data set was also used by \citet{2021A&A...655A.113M}, although their choice of imaging parameters results in a beam width of only 1.75''\,$\times$\,1.91''. However, based on a comparison to H$\alpha$ observations, the authors instead chose a PSF with a width of  1.95''\,$\times$\,2.03''.  
For simplicity of this study, the width of the beam (i.e. the PSF) is set to 2'' at a wavelength of $\lambda = 3.0$\,mm and is then linearly scaled as a function of the observing wavelength $\lambda$.
The considered wavelength range includes the receiver bands that are already available for solar observations with ALMA (bands 3, 5 and 6, i.e. 2.59 - 3.57, 1.42 - 1.90 and 1.09 - 1.42~mm) but in addition also bands 1,2,4, and 7-10, thus covering the whole range from 0.3~mm to 8.5~mm that might be offered in the future. 
The same wavelength-dependent Gaussian kernels are used for degrading the millimetre brightness temperature maps and the H$\alpha$ intensity maps. 
Please note that all three considered snapshots (with 5 minutes solar time between snapshots) are synthesised and then degraded with the different beams. 
In addition, corresponding time-averaged ALMA maps are calculated so that they can be compared to the observational data used by \citet{2019ApJ...881...99M}.

\subsection{Semi-empirical models}

For comparison, the 1D semi-empirical reference models VAL~C \citep{1981ApJS...45..635V} and FAL~C \citep{1993ApJ...406..319F} are used, which are both representative of quiet Sun conditions. Corresponding H$\alpha$ line profiles and mm continuum intensities for these 1D semi-empirical atmosphere models are calculated with the radiative code RH \citep{2001ApJ...557..389U,2015A&A...574A...3P}.

\section{Results}
 \label{Sec:Results}

\subsection{Dependence on spatial resolution}
\label{sec:res_spatres}

\begin{figure*}[ht!]
\centering
\vspace*{-15mm}
\includegraphics[width=1.0\textwidth]{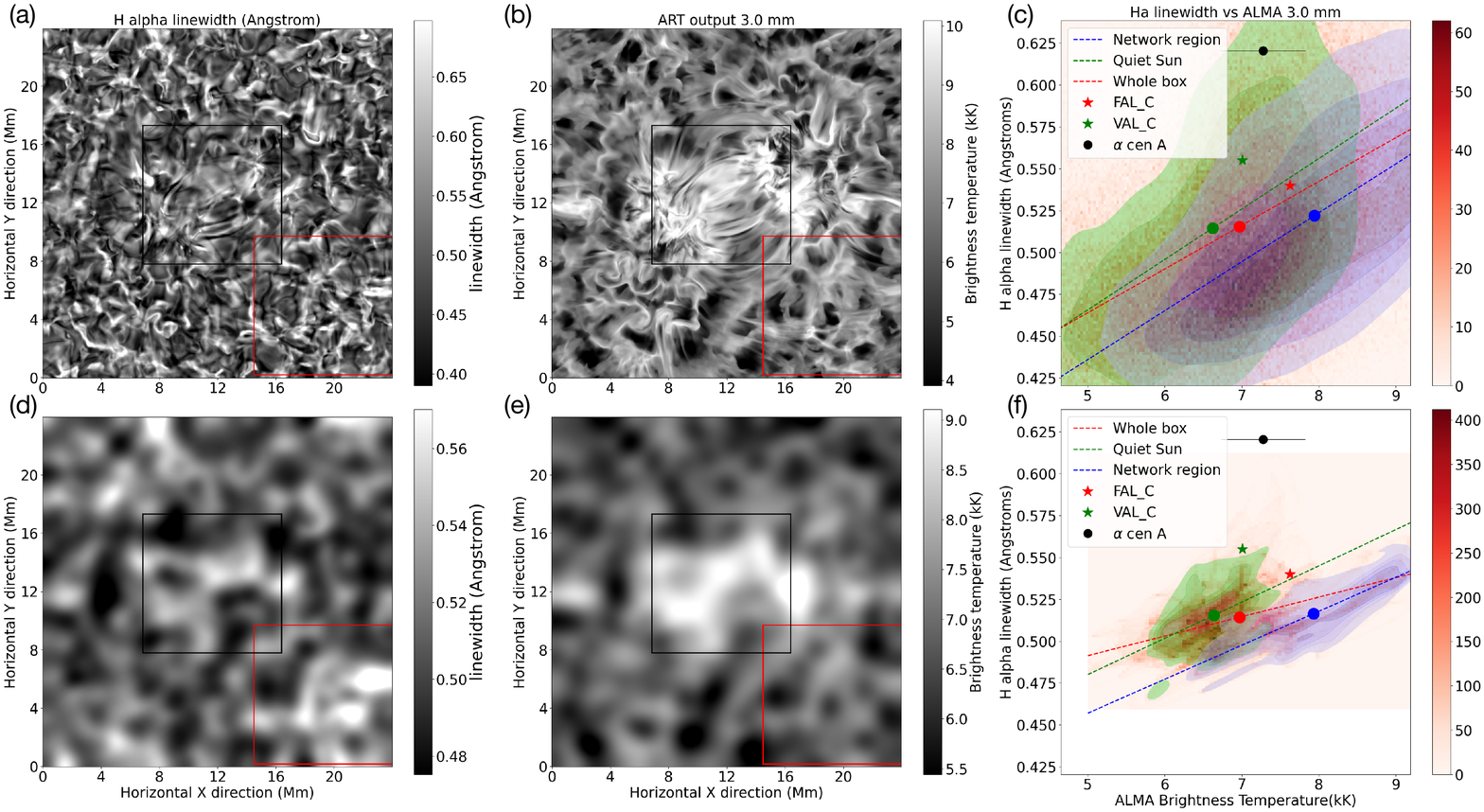}
\vspace*{-20mm}
\caption{Comparisons of the  H$\alpha$ line core width (a,d) and ALMA at 3.0\,mm brightness temperature maps (b,e) at original (top row) and degraded (bottom row) resolution. The black and red boxes on these maps denote the chosen EN and QS regions respectively. The last column is a correlation contour plot between the H$\alpha$ linewidth and ALMA brightness temperature at 3.0\,mm, with contours with linear fits (dashed lines) and means (circles) of the three cases: the whole simulation box in red, the QS box in green and the EN region box in blue. The red and green star data points are for the FAL~C and VAL~C 1D semi-empirical models and the black circle with error bars is an observational data point for G2V type star $\alpha$ Cen A with linewidth 0.62 $\AA$.}
\label{fig:H_a_ALMA_3mm_comparison_leenaarts_formula}
\end{figure*}

Figure~\ref{fig:H_a_ALMA_3mm_comparison_leenaarts_formula} shows the comparison between the EN region and QS region for the original and the degraded resolution corresponding to the assumed ALMA resolution at a wavelength of 3\,mm. The calculated H$\alpha$ line core width for the whole box (panel~a) is very similar to the results in Fig.~9a in \citet{2012ApJ...749..136L}. The H$\alpha$ line core width from the simulation \citep[Fig.~9a in][]{2012ApJ...749..136L} appears to be  formed at a lower height in the chromosphere in comparison to the observation \citep[Fig.~16b in][]{2012ApJ...749..136L}. As the employed wavelength points are very close to the line core, one should expect  to see chromospheric features in the line core width map. In contrast, however, the continuum intensity for 3~mm at original resolution as shown in Fig.~\ref{fig:H_a_ALMA_3mm_comparison_leenaarts_formula}b differs notably from the line core width shown in Fig.~\ref{fig:H_a_ALMA_3mm_comparison_leenaarts_formula}a.  

The maps appear more similar at reduced resolution after convolution with a circular Gaussian kernel corresponding to the spatial resolution at the wavelength of 3~mm (see panels~d and~e).
The black and red boxes on these maps denote the chosen EN and QS regions for further comparison. The EN region is chosen to be centred at the EN as seen in the degraded resolution, and the QS region is placed at the bottom right corner of the simulation box. Both boxes are of the same size of 9.5\,Mm~$\times$~9.5\,Mm (200\,pixels~$\times$~200\,pixels).

The panels c and f of Fig.~\ref{fig:H_a_ALMA_3mm_comparison_leenaarts_formula} show the scatter contour plots for the three sets of pixels: the whole simulation box, the QS box (red square in panels a,b,d,e), and the EN box (black square in panels a,b,d,e). The distribution for the whole simulation box is shown in red contours with the colour bar for the number of pixels falling into the individual bins. The QS and EN distributions are represented by green and blue contours, respectively. The distributions at the original resolution are substantially more spread  
than the distributions at degraded resolution because of the associated loss of variations on small spatial scales. 
This effect is already clear from a visual comparison of the H$\alpha$ linewidth maps in panels a and d and likewise for the mm brightness temperature maps in panels b and e, respectively. 
The red, green and blue points in panels c and f represent the average values for the whole box, QS and EN, respectively.
The impact of applying the ALMA beam can be seen from the spatial power spectral density plots in  Fig.~\ref{fig:PSD_t_avg}.
As seen in the plots for 3.0~mm and 0.8~mm, the power spectral density (PSD) falls drastically for small  spatial scales that are not resolved with respective PSF (beam). 
Another notable observation is that both definitions of linewidths show similar values of power at larger scales. At smaller scales the mm PSD is close to the PSD of the line core width definition, implying better correlations (see  Table~\ref{table:all_three_combined}). Also, the observational 3~mm data shows a very similar trend in PSD to the mm spatially degraded data as expected.

In addition, the H$\alpha$ linewidth and the $T_\mathrm{b}$ at the mm wavelengths for the VAL~C and FAL~C semi-empirical model atmospheres are plotted for comparison. 
The VAL~C model has consistently lower brightness temperature than the FAL~C model as seen in Fig.~\ref{fig:mm_temperatures}. 
When plotted on the linewidth vs temperature planes, these quiet sun models lie in between the EN and QS distributions, demonstrating the agreement between the data from our study and the semi-empirical models.
Applying the beam changes the mean $T_\mathrm{b}$ value for the QS and EN pixel sets as shown in Table~\ref{table:all_three_combined}. The mean EN $T_\mathrm{b}$ decrease and mean QS $T_\mathrm{b}$ increase leave  the mean $T_\mathrm{b}$ of the whole box unchanged. This change in the mean and standard deviation of the brightness temperatures is further shown in Fig.~\ref{fig:mm_temperatures} and discussed in Sect.~\ref{sec:effect_of_spatial_resolution}. The $T_\mathrm{b}$ at 3~mm for the VAL~C and FAL~C models are calculated as 7009\,K and 7624\,K, respectively, which are thus similar to the corresponding average brightness temperature of 6968\,K across the whole horizontal extent of the here employed 3D~model 
(see Table~\ref{table:all_three_combined}).
The value for FAL~C agrees with the calculations reported by 
\citet{2004A&A...419..747L}.
The differences are further discussed in comparison to observations in Sect.~\ref{sec:comparison_with_observations}
It should also be noted that the average QS \mbox{$T_\mathrm{b} = 6621$\,K} found for the 3D model is lower than the reference value of 7300\,K suggested by \citet{2017SoPh..292...88W} for Band~3 ($\lambda = 3.0$\,mm) at the solar disk-centre.
Please refer to Sect.~\ref{sec:comparison_with_observations} for further discussion. 

\begin{figure*}[ht!]
\centering
\vspace*{-15mm}
\includegraphics[width=1.0\textwidth]{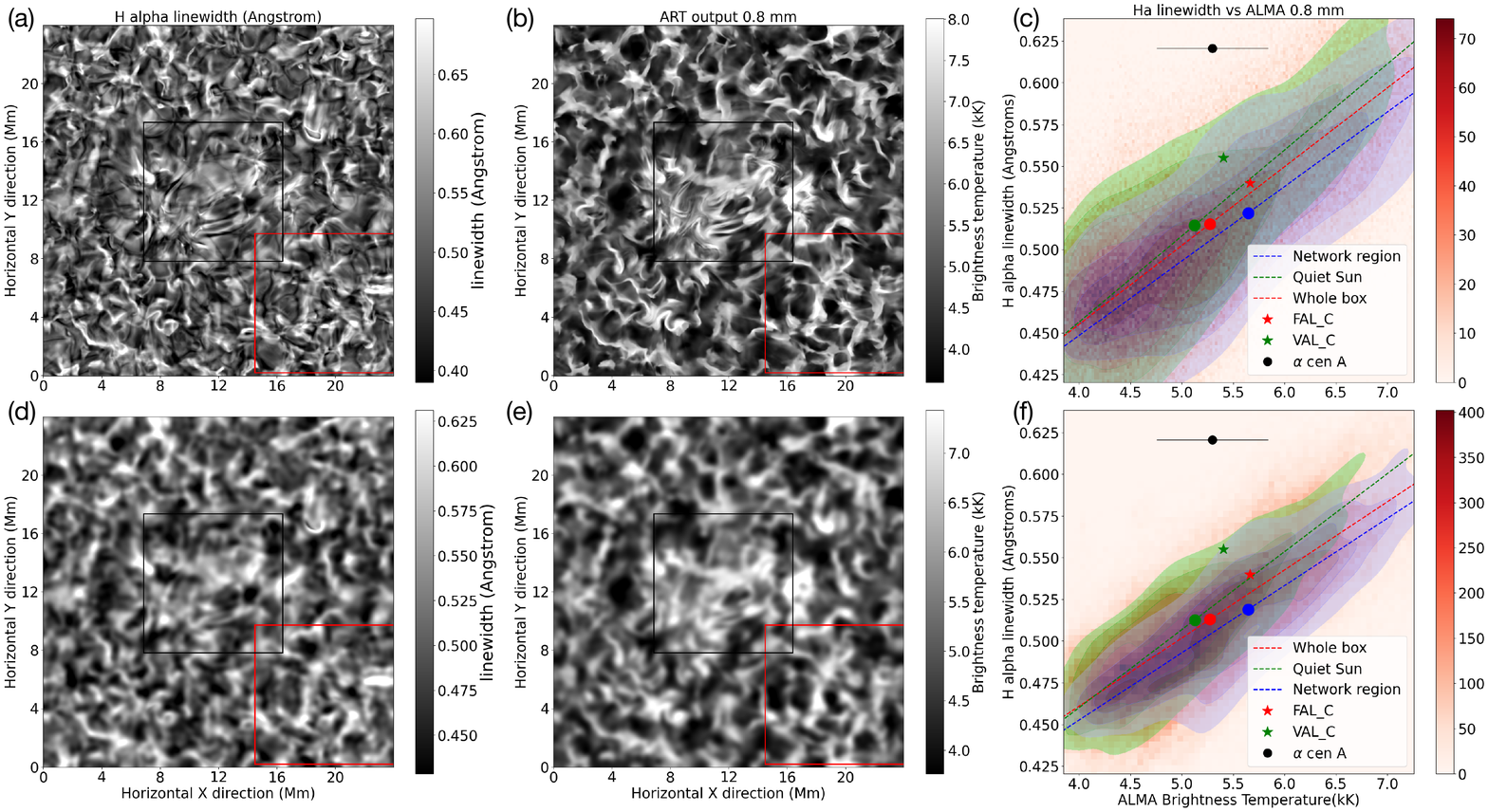}
\vspace*{-20mm}
\caption{Same as Fig.~\ref{fig:H_a_ALMA_3mm_comparison_leenaarts_formula} but for a wavelength of 0.8~mm.}
\label{fig:H_a_ALMA_0.8mm_comparison_leenaarts_formula}
\end{figure*}

\begin{figure*}[t!]
\includegraphics[width=0.5\textwidth]{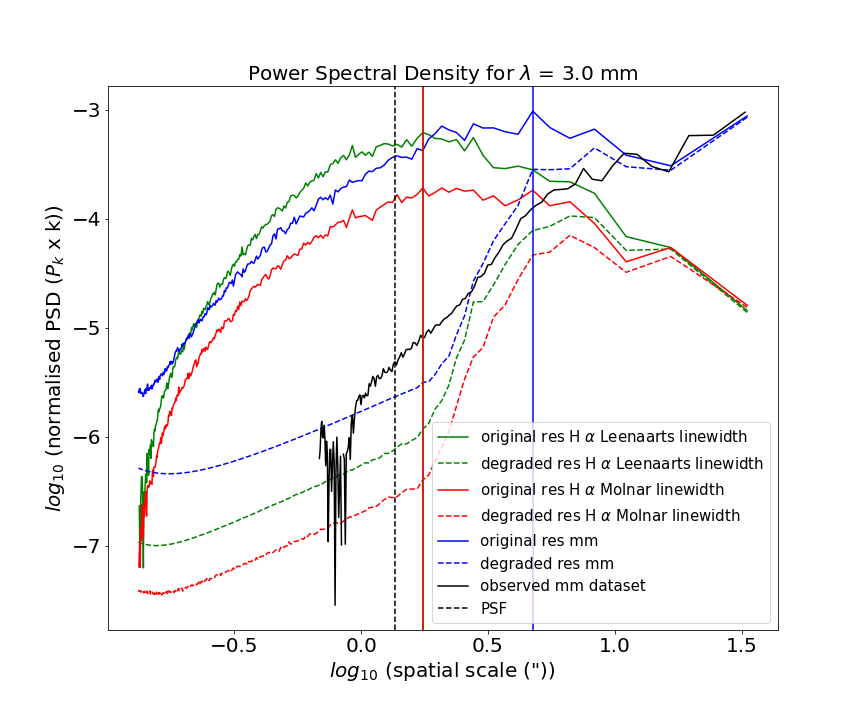}
\includegraphics[width=0.5\textwidth]{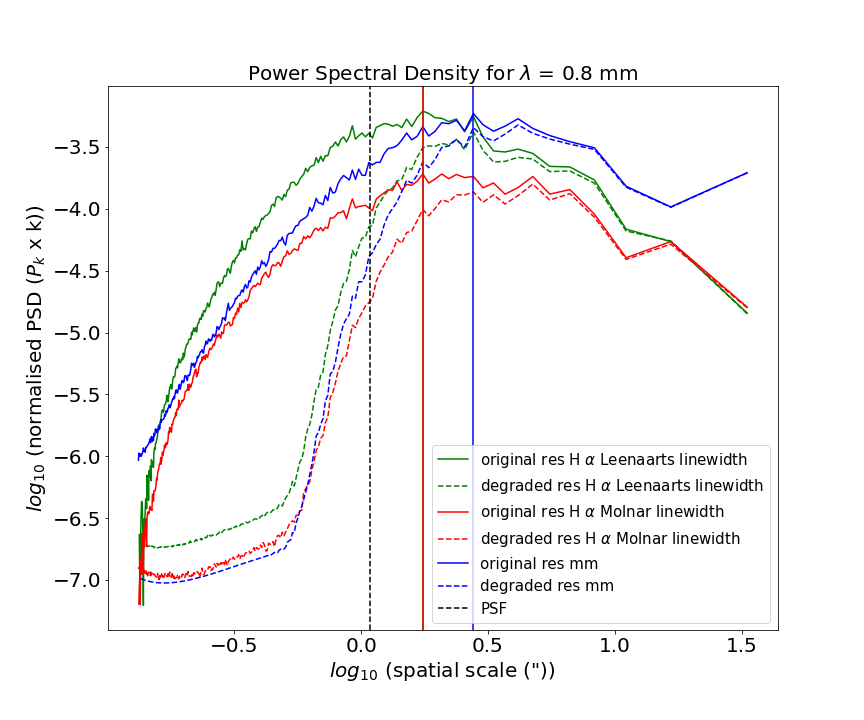}
\vspace*{-8mm}
\caption{Normalised power spectral density $kP_k$ plots for time-averaged datasets of mm brightness temperatures, and both definitions of linewidths at 3.0~mm and 0.8~mm. The solid lines represent the original resolution data and the dashed lines, the spatially degraded data, green for \citep{2012ApJ...749..136L} linewidth, red for  \citet{2019ApJ...881...99M} definition of linewidth and blue for the corresponding mm wavelength. The vertical lines represent the dominant spatial scales (in arcsec) for the original resolution, for \citep{2012ApJ...749..136L} definition of linewidth and  \citet{2019ApJ...881...99M} definition of linewidth respectively. The black dashed line represents the PSF corresponding to the mm wavelength, showing the resolution limit for the degraded dataset. 
In the left panel, the spatial power spectral density for the Band~3 ALMA observations (ADS/NRAO.ALMA\#2016.1.01129.S) as used by \citet{2019ApJ...881...99M} and shown in Fig.~\ref{fig:Molnar} are plotted as the black solid line. The power spectral density was calculated for all time steps in the time range considered by \citet{2019ApJ...881...99M} and then averaged in time.}
\label{fig:PSD_t_avg}
\end{figure*}

\begin{figure}[t!]
\vspace*{-5mm}
\centering
\includegraphics[width=0.49\textwidth]{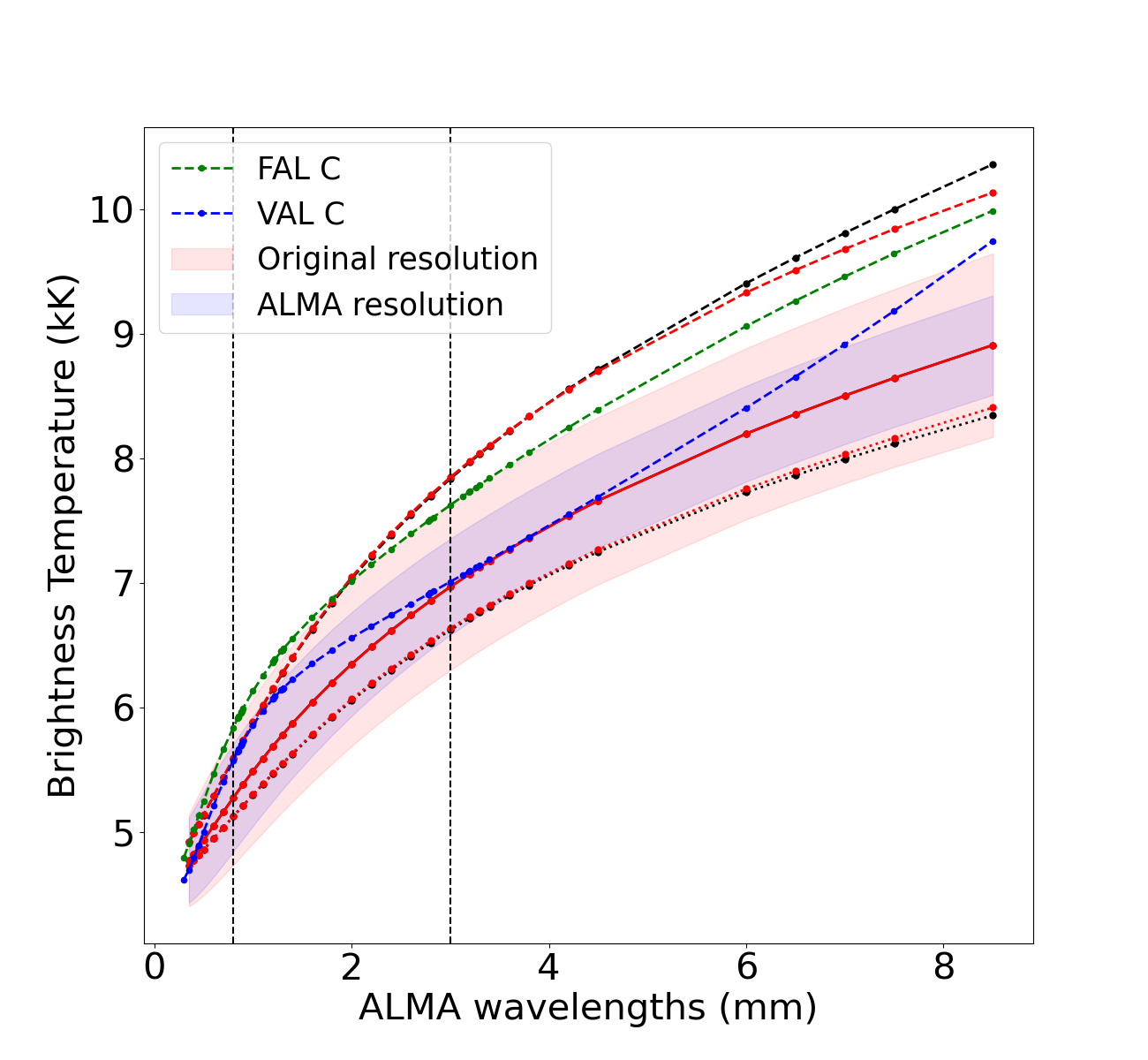}
\vspace*{-5mm}
\caption{Comparison of the mean brightness temperatures for mm wavelengths at the resolution of the original simulations (in black) and the resolution of the observations from the ALMA observatory (in red). In both these cases the whole simulation box (solid lines), the QS box(dotted lines) and the EN box (dashed lines) are taken separately. The pink and blue shaded regions show one standard deviation region around the mean for the original resolution and the degraded resolution case respectively. The green and blue dashed lines show brightness temperature trends for FAL~C and VAL~C models respectively. The black dashed vertical lines mark 0.8~mm and 3~mm.}
\label{fig:mm_temperatures}
\end{figure}

\begin{figure}
\vspace*{-5mm}
\centering
\includegraphics[width=0.5\textwidth]{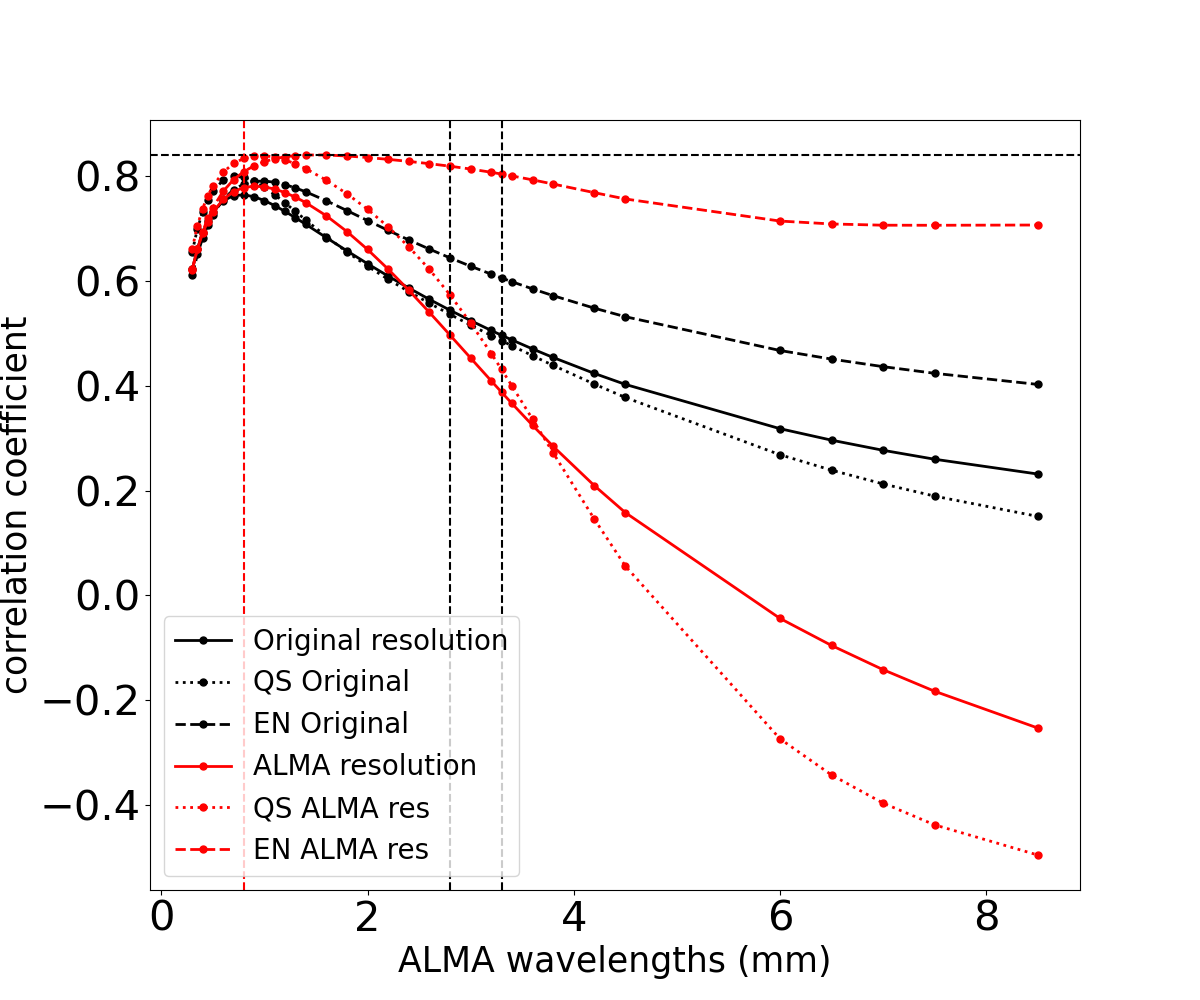}
\caption{Comparison of the calculated Pearson correlation coefficients with two different resolutions. Black vertical lines show the ALMA band 3 range and black horizontal lines show the Molnar et. al. correlation coefficient = 0.84. and the red vertical line shows the maximum correlation in the original resolution case which is at 0.8\,mm.}
\label{fig:H_a_ALMA_mm_correlations_Leenaarts_et_al}
\end{figure}

\begin{figure}
\vspace*{-5mm}
\centering
\includegraphics[width=0.5\textwidth]{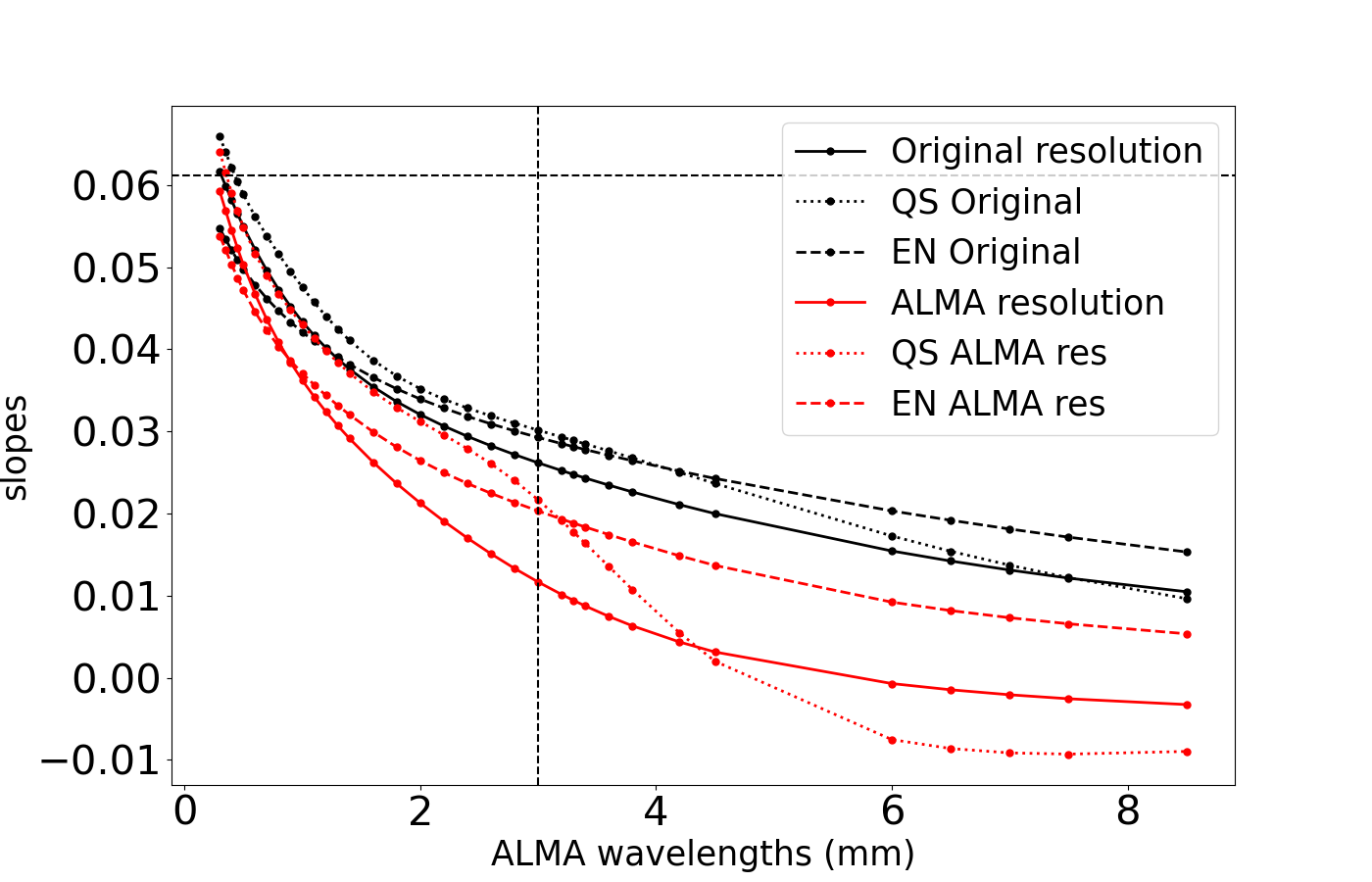}
\caption{Comparison of the calculated slopes for the scatter plots, legend same as Fig.~\ref{fig:mm_temperatures}. The black lines show the Molnar et. al. slope = 0.0612 at 3\,mm.}
\label{fig:H_a_ALMA_mm_slopes_Leenaarts_et_al}
\end{figure}

\begin{figure*}[ht!]
\centering
\vspace*{-15mm}
\includegraphics[width=1.0\textwidth]{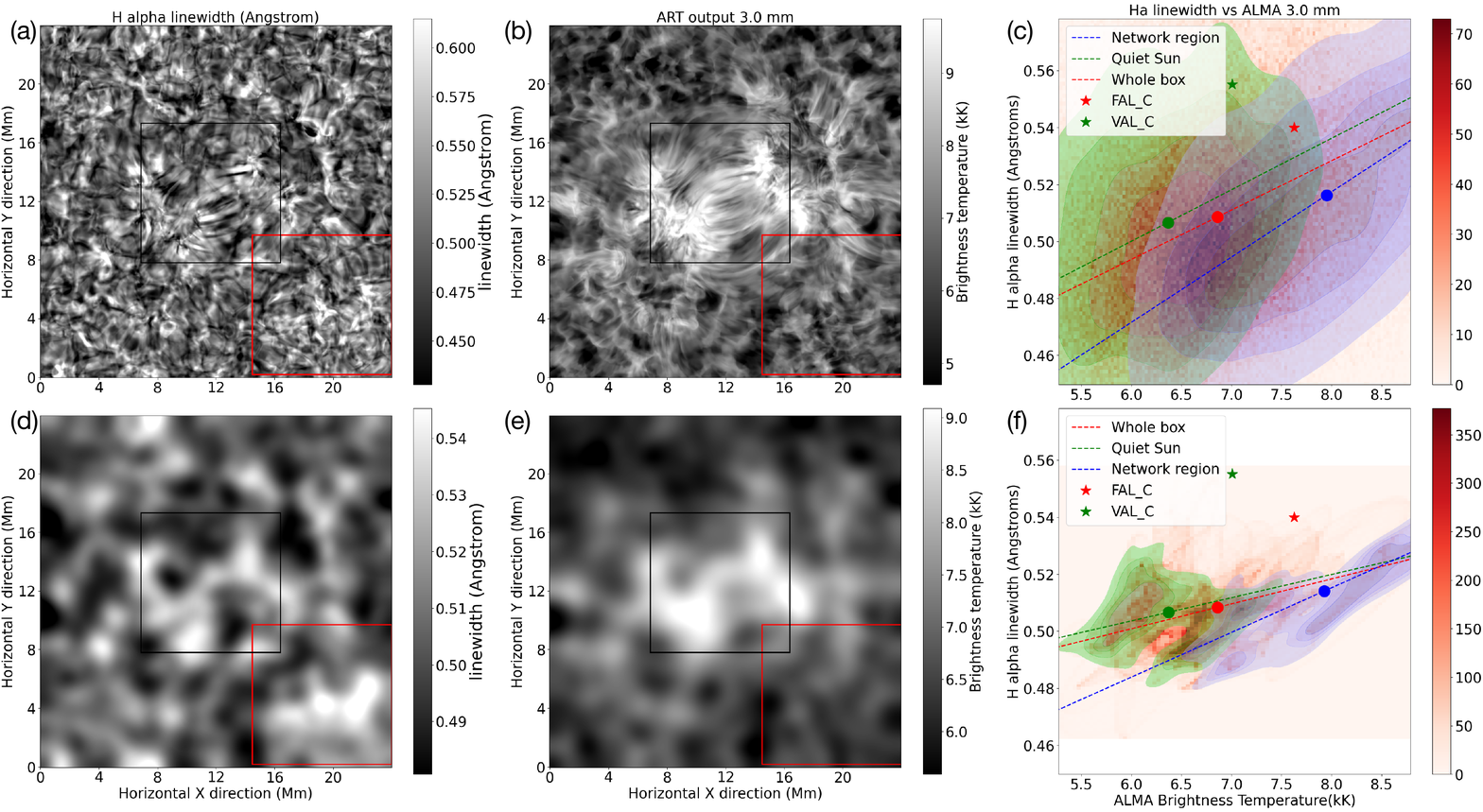}
\vspace*{-20mm}
\caption{Same as Fig.~\ref{fig:H_a_ALMA_3mm_comparison_leenaarts_formula}, for time-averaged dataset.}
\label{fig:H_a_ALMA_3mm_comparison_leenaarts_formula_time_averaged}
\end{figure*}

\begin{figure*}[ht!]
\centering
\vspace*{-15mm}
\includegraphics[width=1.0\textwidth]{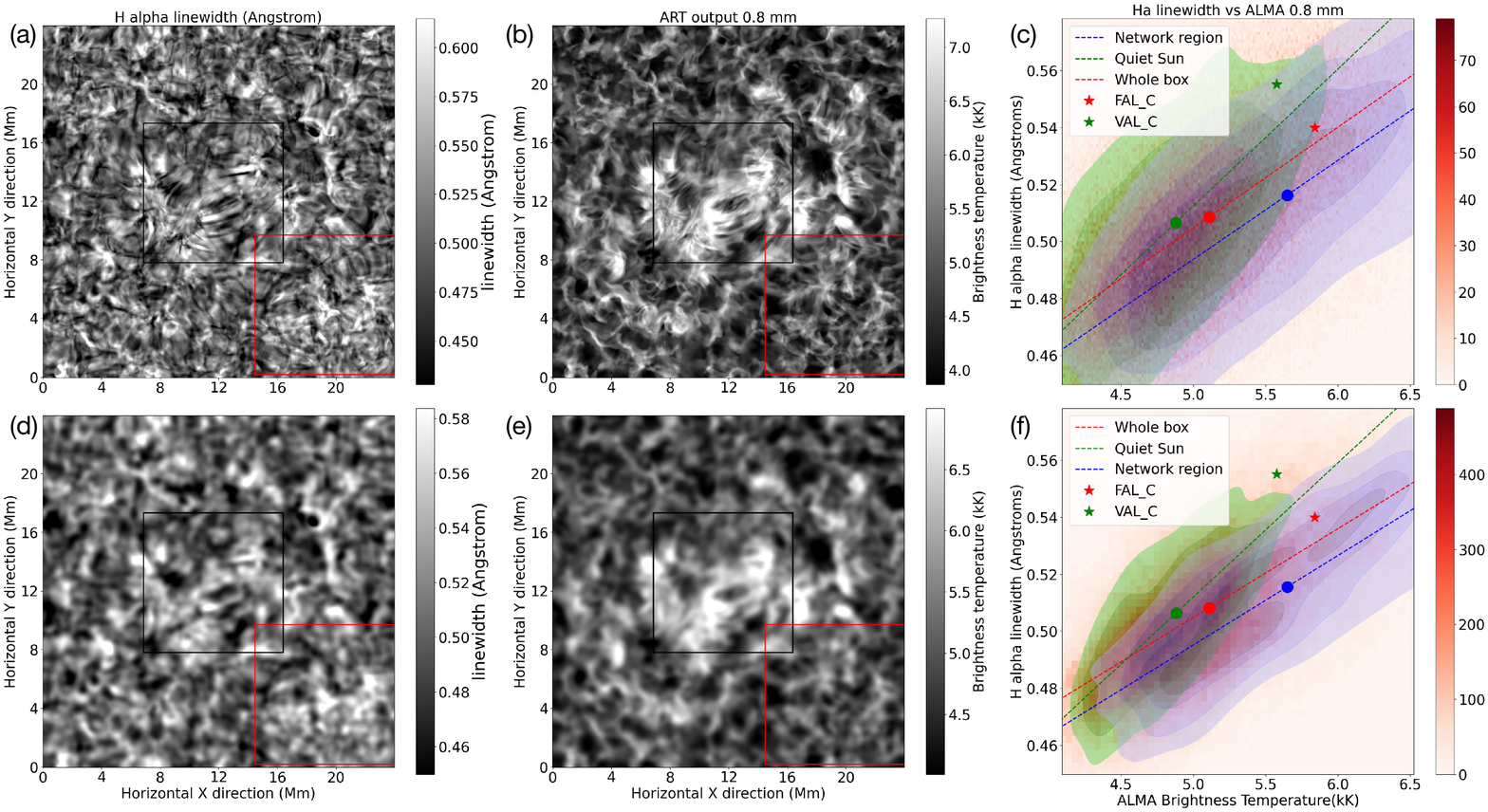}
\vspace*{-20mm}
\caption{Same as Fig.~\ref{fig:H_a_ALMA_0.8mm_comparison_leenaarts_formula}, for time-averaged dataset.}
\label{fig:H_a_ALMA_0.8mm_comparison_leenaarts_formula_time_averaged}
\end{figure*}

\begin{table*}[ht!]
\begin{center}
\begin{tabular}{|c|c|c|c|c|c|c|c|}
\hline
\multicolumn{2}{|c}{}& 
\multicolumn{3}{|c}{Original resolution}& 
\multicolumn{3}{|c|}{Reduced resolution}\\
\cline{2-8}
 & wavelength  & Whole box  & 	 QS	 &  EN & Whole box  &  QS &  EN \\
\hline
\hline
\multicolumn{2}{|c}{Single snapshot:} & 
\multicolumn{2}{c}{line core width calculated with} &
\multicolumn{2}{c} {\citep{2012ApJ...749..136L} formula.} &
\multicolumn{2}{c|}{ } \\		
\hline
Temperatures &	0.8~mm &	5.275	 &5.124 &	5.590 &	5.275 &	5.126 &	5.593	\\
(kK) &	3.0~mm &	6.968 &	6.621	 &7.835 &	6.968 &	6.637 &	7.848	\\
\hline
Correlation &	0.8~mm &	0.764	&0.798&	0.785	&0.778&	0.834&	0.808	\\
 coefficients &	3.0~mm &	0.524	&0.516&	0.628	&0.453&	0.520	&0.813\\

\hline
Slopes &	0.8~mm &	0.047 & 0.052&	0.045&	0.041&	0.047	&0.040	\\
&	3.0~mm & 0.026&	0.030&	0.029&	0.012&	0.022&	0.020	\\
\hline
\hline
\multicolumn{2}{|c}{Time-averaged:} & 
\multicolumn{2}{c}{line core width calculated with} &
\multicolumn{2}{c} {\citep{2012ApJ...749..136L} formula.} &
\multicolumn{2}{c|}{ } \\		
\hline
Temperatures &	0.8~mm &		5.111&	4.881&	5.609	&5.111	&4.881&	5.609	\\
(kK) &	3.0~mm &	6.862&	6.367&	7.891	&6.862	&6.371&	7.878	\\
\hline
Correlation &	0.8~mm &	0.668	& 0.705& 	0.712	& 0.696	& 0.751& 	0.755	\\
 coefficients &	3.0~mm &	0.461& 	0.398& 	0.583	& 0.470	& 0.316& 	0.755\\

\hline
Slopes &	0.8~mm &	0.035	& 0.048& 	0.035& 	0.031& 	0.047	& 0.031	\\
&	3.0~mm & 0.017	& 0.018	& 0.023	& 0.009& 	0.008	& 0.016	\\
\hline
\hline
\multicolumn{2}{|c}{Single snapshot:} & 
\multicolumn{2}{c}{linewidth calculated with} &
\multicolumn{2}{c} { \citet{2019ApJ...881...99M} formula.} &
\multicolumn{2}{c|}{ } \\						
\hline

Correlation &	0.8~mm &	0.733 &	0.746 &	0.797 &	0.770	 &0.813 &	0.854	\\
 coefficients &	3.0~mm &	0.443 &	0.443 &	0.545 &	0.384 &	0.411 &	0.819\\

\hline
Slopes &	0.8~mm &	0.042 &	0.043 &	0.040 &	0.044 &	0.048 &	0.041	\\
&	3.0~mm &	0.020	 &0.023	 &0.022	 &0.014 &	0.022	 &0.022	\\
\hline
\hline
\multicolumn{2}{|c}{Time-averaged:} & 
\multicolumn{2}{c}{linewidth calculated with} &
\multicolumn{2}{c} { \citet{2019ApJ...881...99M} formula.} &
\multicolumn{2}{c|}{ } \\		
\hline

Correlation &	0.8~mm &0.729&	0.692&	0.837&	0.771&	0.724&	0.894	\\
 coefficients &	3.0~mm &	0.479&	0.343&	0.609&	0.610&	0.294&	0.829\\

\hline
Slopes &	0.8~mm &	0.040&	0.045&	0.043&	0.039&	0.046&	0.043	\\
&	3.0~mm &	0.019&	0.015& 0.025&	0.015& 0.008&	0.025	\\
\hline
\end{tabular}
\caption{Brightness temperatures in kK, Correlation coefficients and Slopes for the three pixel sets (whole box, QS and EN) each for two resolutions (original and degraded) for the single snapshot and time-averaged dataset, with both the definitions used for calculating the linewidth.}
\vspace{-8mm}
\label{table:all_three_combined}
\end{center}
\end{table*}

The wavelength of 0.8~mm is of particular interest for this study, too, as will be discussed in Sects.~\ref{sec:correlations_and_slopes} and \ref{sec:comparison_with_observations}.
{The results for $\lambda = 0.8$\,mm are presented in Fig.~\ref{fig:H_a_ALMA_0.8mm_comparison_leenaarts_formula}, which corresponds to the above-discussed Fig.~\ref{fig:H_a_ALMA_3mm_comparison_leenaarts_formula} for $\lambda = 3.0$\,mm. 
While the H$\alpha$ line core width map at the original resolution (panel~a) is the same in both figures, the mm brightness temperature maps (panel~b) and the degraded maps  (panels d and e) differ. The H$\alpha$ line core width map 
(Fig.~\ref{fig:H_a_ALMA_0.8mm_comparison_leenaarts_formula}a) and the mm continuum map at 0.8\,mm (Fig.~\ref{fig:H_a_ALMA_0.8mm_comparison_leenaarts_formula}b)
are already very similar at the original resolution and even more so at the lower resolution as shown in Fig.~\ref{fig:H_a_ALMA_0.8mm_comparison_leenaarts_formula}d-e. The PSF at 0.8\,mm is smaller in width compared to the 3.0\,mm PSF by a factor of $0.8/3.0$, which can be seen from a visual comparison of Fig.~\ref{fig:H_a_ALMA_3mm_comparison_leenaarts_formula}e and Fig.~\ref{fig:H_a_ALMA_0.8mm_comparison_leenaarts_formula}e. 
The 3.0\,mm maps appear more blurred, mainly as a result of the wider PSF. 
The same effect can be seen for the H$\alpha$ line width maps, which were derived from the  H$\alpha$ line intensity maps after these were all degraded accordingly for each wavelength point.

The slopes in the EN and QS cases are very similar to each other and to that for the whole box as seen in Table~\ref{table:all_three_combined}. In contrast to Fig.~\ref{fig:H_a_ALMA_3mm_comparison_leenaarts_formula}c,f, the scatter plots in Fig.~\ref{fig:H_a_ALMA_0.8mm_comparison_leenaarts_formula}c,f show that the slopes for the QS and EN distributions (0.052 and 0.045) are very similar to the values for the maps at degraded resolution. On the other hand, for 3~mm, the values are 0.03 and 0.029, which show a higher departure from each other. The mean $T_b$ values for the QS region and the whole box are quite close as seen in Table~\ref{table:all_three_combined}, as the atmosphere is mainly QS. The VAL~C and FAL~C data points are closer to the mean $T_b$ for the EN region, possibly because of the reasons discussed before.
Treating the synthetic mm and the H$\alpha$ linewidth map with the generated PSFs for corresponding wavelengths (see Appendix~\ref{Appendix_B} for representative plots similar to Figs.~\ref{fig:H_a_ALMA_3mm_comparison_leenaarts_formula},~\ref{fig:H_a_ALMA_0.8mm_comparison_leenaarts_formula} for ALMA Bands), the scatter and the linear fits were plotted and the slopes and the correlation coefficients were calculated. 
The trends in correlations and slopes for all considered wavelengths are presented in Sect.~\ref{sec:correlations_and_slopes} and shown in Figs.~\ref{fig:H_a_ALMA_mm_correlations_Leenaarts_et_al} and \ref{fig:H_a_ALMA_mm_slopes_Leenaarts_et_al}, respectively.

\subsection{Correlations and slopes}
\label{sec:correlations_and_slopes}

The correlation of the H$\alpha$ linewidth maps and the mm brightness temperature maps and the dependence of this correlation on millimetre wavelength and thus spatial resolution are quantified through the Pearson correlation coefficient between the H$\alpha$ linewidth maps and mm brightness temperature maps and linear fits the distributions as shown, for instance, in  Figs.~\ref{fig:H_a_ALMA_3mm_comparison_leenaarts_formula},~\ref{fig:H_a_ALMA_0.8mm_comparison_leenaarts_formula}. The slopes, as well as the intercept of the linear fits, are calculated by linear regression for the whole box, and the EN and QS regions separately.

The resulting correlation coefficients as a function of wavelength are shown in Fig.~\ref{fig:H_a_ALMA_mm_correlations_Leenaarts_et_al} at original resolution (black lines) and after degradation with the synthetic ALMA PSFs (red lines) corresponding to the wavelength. 
It should be noted that the mm brightness temperature maps exhibit differences as a function of wavelength due to the different formation height ranges at a given wavelength. Although the formation heights can vary notably across the model, they nonetheless increase on average with increasing wavelength. Consequently, the mm brightness temperature maps at original resolution show a maximum correlation with the H$\alpha$ linewidth maps at a given wavelength as can be seen most clearly from the peak for the QS correlation (black dotted line) for wavelengths in the range from 0.5\,mm to 1.8\,mm. Peaks are also present for the whole box and the EN region with a decrease towards shorter and longer wavelengths in all cases. 

Reducing the spatial resolution has a notable impact on the correlation coefficients (see Fig.~\ref{fig:H_a_ALMA_mm_correlations_Leenaarts_et_al}), 
resulting in generally higher values as compared to the correlation at the original resolution.  
The QS region correlations are lower than for the whole box in both cases and the EN region correlations are slightly higher in the degraded resolution case and for most wavelengths, for the original resolution.
While also the correlation for the degraded resolution maps exhibits peaks, there is a systematic increase in the correlation between the degraded resolution maps as a function of wavelength for longer wavelengths.
This effect can be attributed primarily to the spatial resolution, i.e., the width of the PSF which increases with wavelength. Consequently, the structure below a corresponding (small) spatial scale is blurred, which affects the resulting correlation coefficient. With increasing wavelength more and more of this atmospheric fine structure is blurred as already discussed in Sect.~\ref{sec:res_spatres} and seen from a comparison of Figs.~\ref{fig:H_a_ALMA_3mm_comparison_leenaarts_formula} ~and~\ref{fig:H_a_ALMA_0.8mm_comparison_leenaarts_formula}
(see Table~\ref{table:all_three_combined} for the correlation coefficients for 3.0\,mm and 0.8\,mm). 
At the longest wavelengths, however, even structure on relatively large spatial scales is lost. Consequently, the data ranges for these very blurred maps become very narrow, which then results in a high correlation coefficient.  
The impact of a lower spatial resolution on the data ranges is also visible in Fig.~\ref{fig:H_a_ALMA_3mm_comparison_leenaarts_formula}c,f and \ref{fig:H_a_ALMA_0.8mm_comparison_leenaarts_formula}c,f: The shown distributions become notably narrower as a function of spatial resolution for the whole box, QS region and EN region, being broadest at the original resolution and narrowest when degraded with the 3.0\,mm PSF.
Most importantly, the highest correlation for the degraded maps with a value of 0.844 is found for the EN region at 1.1\,mm, whereas the QS region has a maximum of 0.839 at 1.0\,mm. The correlation for the whole box peaks with 0.799 at a wavelength of 1.0\,mm, too.  
These results suggest that (i)~the highest correlation between the H$\alpha$ line core width and mm continuum brightness temperature maps is found for wavelengths at 1.0\,mm, which corresponds to ALMA Band~7, and that (ii)~the correlation between observed H$\alpha$ line core width maps and co-simultaneously obtained ALMA Band~7 brightness temperature maps is expected to be higher than what was found by \citet{2019ApJ...881...99M} for Band~3 observations (at 3.0\,mm). See Sect.~\ref{sec:comparison_with_observations} for further discussion.

The blue contours in Fig.~\ref{fig:H_a_ALMA_0.8mm_comparison_leenaarts_formula}c,f, which represent the EN region, show higher H$\alpha$ line core widths and mm brightness temperatures. On the other hand, the green contours, for QS regions, show lower intensities in mm and lower H$\alpha$ linewidths, in both the resolutions, more distinct in the degraded resolution case. As the smaller network-internetwork features get blurred, the higher (lower) intensities in the QS (EN) region get blurred with neighbouring lower (higher) intensities and hence the green (blue) contours shift downwards (upwards). 
The slopes of the fitted lines for the three distributions change with resolution. The slope decreases for the EN case but increases for the QS case with the degraded resolution as can be seen in Fig.~\ref{fig:H_a_ALMA_mm_slopes_Leenaarts_et_al} for all the cases in the dataset. 

The slopes show a general decreasing trend with wavelength. As the wavelength increases, the radiation originates from a slightly higher layer of the atmosphere which is on average slightly hotter. 
As seen in Fig.~\ref{fig:mm_temperatures}, the temperatures at different wavelengths are also dependent on the formation heights, in addition to the resolution. This is evident from the decreasing slopes shown in Fig.~\ref{fig:H_a_ALMA_mm_slopes_Leenaarts_et_al}. For the degraded resolution case, the difference in the slope for QS and EN regions is more than the original resolution case.
In particular, for the EN case, as seen in Figs.~\ref{fig:H_a_ALMA_3mm_comparison_leenaarts_formula} and \ref{fig:H_a_ALMA_0.8mm_comparison_leenaarts_formula}, in the degraded resolution case (panels d and e), the brightness temperatures decrease, 
decreasing slopes significantly. For the QS case, on the other hand, the slopes remain comparable, or approximately the same in both resolutions. This is explained by the 
blurring 
effect due to the application of the ALMA PSFs, but in the QS region, the features are more equally distributed.  
Hence the slopes do not
significantly change with the resolution in the millimetre range of interest (Bands 3-6).
The line widths derived from the time-averaged H$\alpha$ intensity maps following the definition by  \citet{2019ApJ...881...99M} produce very similar trends and also indicate the best match to occur at wavelengths corresponding to ALMA Band~7. 
Please refer to Sects.~\ref{sec:temp_res},~\ref{sec:comparison_with_observations} for further discussion.

\subsection{Dependence on temporal resolution}
\label{sec:temp_res}

Figures~\ref{fig:H_a_ALMA_3mm_comparison_leenaarts_formula_time_averaged} and \ref{fig:H_a_ALMA_0.8mm_comparison_leenaarts_formula_time_averaged} show the original and ALMA resolution maps for the dataset averaged over 10 minutes. The comparison of features in Figs.~\ref{fig:H_a_ALMA_3mm_comparison_leenaarts_formula_time_averaged}a,b,~\ref{fig:H_a_ALMA_3mm_comparison_leenaarts_formula}a,b and Figs.~\ref{fig:H_a_ALMA_0.8mm_comparison_leenaarts_formula_time_averaged}a,b,~\ref{fig:H_a_ALMA_0.8mm_comparison_leenaarts_formula}a,b respectively, shows that the time averaging causes the features to smooth out, as expected. The QS, EN and whole box contours show less spread in the time-averaged case than in the single snapshot case as observed in Figs.~\ref{fig:H_a_ALMA_3mm_comparison_leenaarts_formula_time_averaged}c,f,~\ref{fig:H_a_ALMA_3mm_comparison_leenaarts_formula}c,f and Figs.~\ref{fig:H_a_ALMA_0.8mm_comparison_leenaarts_formula_time_averaged}c,f,~\ref{fig:H_a_ALMA_0.8mm_comparison_leenaarts_formula}c,f respectively.
When calculating the H$\alpha$ line core width from the time-averaged maps and comparing it to the time-averaged mm brightness temperature maps, the correlation between the H$\alpha$ line core width maps and the mm  brightness temperature maps at original resolution decreases, whereas the correlation increases significantly for the corresponding maps at reduced (i.e. ALMA) resolution. 
The correlation is highest for the EN case and the lowest for the QS case at both resolutions. This can be attributed to the differences in the small-scale features in the QS region, which get averaged over time and hence do not match in the two data sets at degraded resolution. Similarly, the slopes for the QS region are also lower than for the whole box, whereas the slopes for the EN region are higher than for the whole box. The slopes do not show very drastic changes from the single snapshot case, but they are monotonically decreasing for all pixel sets.

\section{Discussion}
\label{Sec:Discussion} 

The H$\alpha$ line core widths correlate best with the mm continuum brightness temperatures at a wavelength of 0.8\,mm. 
The Pearson correlation coefficients for the degraded maps are determined as 0.778 for the whole box,  0.808 for the QS region, and 0.834 for the EN region (see Table~\ref{table:all_three_combined}). 
These results imply that, despite the similarities of the H$\alpha$ linewidth with the observational ALMA Band~3 data found by \citet{2019ApJ...881...99M}, a closer match with Band~7 is to be expected. 
The immediate next step therefore would be to compare observational ALMA Band~7 data with co-observed H$\alpha$ data but, to the best knowledge of the authors, no such data is currently available for the study presented here.
 
\subsection{Effect of spatial resolution}
\label{sec:effect_of_spatial_resolution}

One of the motivations of this study is to use high-resolution synthetic data at mm wavelengths to investigate how they can complement other diagnostics, like H$\alpha$, for determining the properties of small-scale features in the solar atmosphere. Even though the resolution of ALMA observational data is outstanding compared to previous observations at millimetre wavelengths, 
the resolution is very low 
compared to what is typically achieved at optical wavelengths \citep[see, e.g.,][and references therein]{2006A&A...456..697W,2016SSRv..200....1W}.

The comparison between maps at the original resolution and maps at the ALMA resolution using Gaussian kernels as simplified PSFs for the respective wavelengths (see Figs.~\ref{fig:H_a_ALMA_3mm_comparison_leenaarts_formula} and \ref{fig:H_a_ALMA_0.8mm_comparison_leenaarts_formula}) shows that there are two effects in action: the change in the formation layer for the mm data, and the decreasing resolution due to the increasing wavelength. 
These two effects cannot be disentangled in observational data, but with a simulation like the one employed here the physical and instrumental effects can be separated. In Fig.~\ref{fig:mm_temperatures}, the effect of the change in formation height for the continuum radiation at mm wavelengths and the change in resolution for the given wavelength are demonstrated. 
Although the average brightness temperature of the whole simulation box remains the same when degrading the resolution at a given wavelength, the width of the brightness temperature distribution (as represented by the shaded areas in the figure) reduces with degrading resolution.
The increase in correlations with the degradation of the resolution can be attributed to this phenomenon. Also, the slopes are affected as the standard deviations in the data with the reduced resolution, which is evident from Fig.~\ref{fig:mm_temperatures}. 

To understand this impact of the spatial resolution more quantitatively, the values of the correlations for all six cases for $\lambda = 0.8$~mm and $\lambda = 3.0$~mm are listed in Table~\ref{table:all_three_combined}. The correlations significantly increase with the degradation of the resolution as observed in Fig.~\ref{fig:H_a_ALMA_mm_correlations_Leenaarts_et_al}. On the other hand, as seen in Fig.~\ref{fig:H_a_ALMA_mm_slopes_Leenaarts_et_al} and Table~\ref{table:all_three_combined}, the slopes do not change significantly. 
Hence, it can be safely speculated that the H$\alpha$ linewidth and mm brightness temperatures 
are correlated and follow the linear fit fairly well.

In Fig.~\ref{fig:formation_heights}, the formation heights for the continuum at wavelengths of 3.2~mm, 1.8~mm and 0.8~mm, along with the vertical temperature slice through the model atmosphere are presented. Two more formation height profiles are plotted for the linewidth definition by  \citet{2019ApJ...881...99M}  \citep[cf.][]{2009A&A...503..577C} and for the line core width as defined by \citet{2012ApJ...749..136L}. In both cases, the plotted line is the average formation height of the wavelength points in the blue and red line wing that result from the calculation of the linewidth. It is observed that the \citet{2012ApJ...749..136L} definition agrees better with the 0.8~mm formation height range. The  \citet{2019ApJ...881...99M} linewidth is rather a diagnostic of the lower atmosphere.

\subsection{Comparison with observations}
\label{sec:comparison_with_observations}

\begin{figure*}[ht!]
\vspace*{-15mm}
\includegraphics[width=1.0\textwidth]{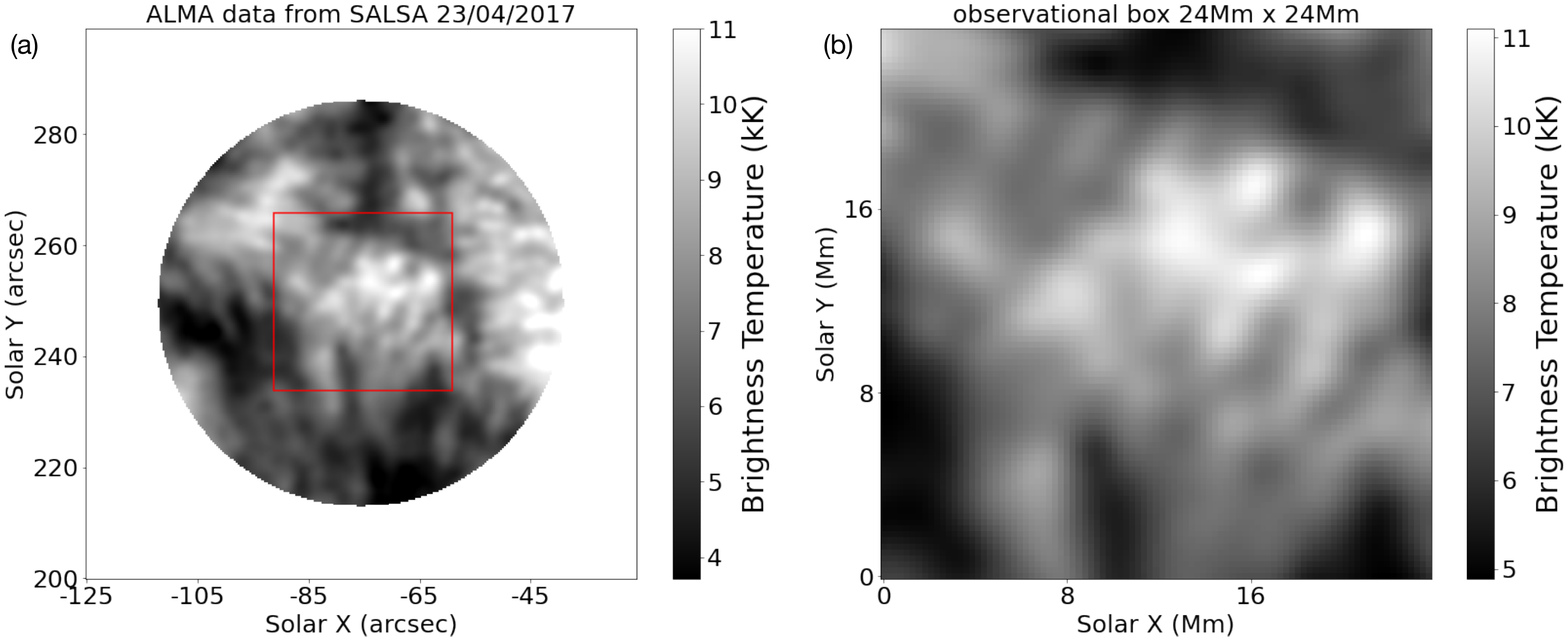}
\vspace*{-45mm}
\caption{
Time-averaged ALMA Band~3 brightness temperature for the same measurement set as used by \citet[][]{2019ApJ...881...99M}, i.e. ADS/NRAO.ALMA\#2016.1.01129.S from 23 April 2017. The data were averaged over the period from 17:31 to 17:41 UTC. (a)~The whole Field of View (FOV). (b)~The central 24\,Mm\,$\times$\,24\,Mm part, also indicated by the red box in panel~a for comparison with the simulated data.}
\label{fig:Molnar}
\end{figure*}

To compare these results with the observational study by \citet{2019ApJ...881...99M}, we used the same definition of the linewidth as used by them which follows the approach by 
\citep[][see Sect.~\ref{sec:deflinewidth}]{2009A&A...503..577C}. 
The corresponding H$\alpha$ and mm maps at the original and degraded resolution with the scatter contour plots are shown in Figs.~\ref{fig:H_a_ALMA_3.0mm_comparison_time_averaged} and \ref{fig:H_a_ALMA_0.8mm_comparison_time_averaged}. In Figs.~\ref{fig:H_a_ALMA_mm_correlations_time_averaged}c,f and \ref{fig:H_a_ALMA_mm_slopes_time_averaged}c,f, the black dashed lines show the respective correlation and slope calculated by \citet{2019ApJ...881...99M} for the observational data obtained with ALMA in Band~3 (3\,mm) on 23 April 2017 (see their Fig.~3). 
A brightness temperature map similar to the one shown in Fig.~3 by \citet{2019ApJ...881...99M} was produced by time-averaging the corresponding data that is publicly 
available on the Solar ALMA Science Archive \citep[SALSA,][]{2022A&A...659A..31H} 
 (see Fig.~\ref{fig:Molnar}).
The differences seen in the data from \citeauthor{2019ApJ...881...99M} and the data taken from SALSA are due to the difference in the applied imaging including the choice of CLEAN parameters and the combination of the interferometric and Total Power (TP) data as \citeauthor{2019ApJ...881...99M} follow the standard approach while the SALSA data is produced with the Solar ALMA Pipeline \citep{2020A&A...635A..71W}. Please refer to \citet{2017SoPh..292...88W} and \citet{2017SoPh..292...87S} for a detailed discussion of the uncertainties of the interferometric and TP data, which both affect the resulting absolute brightness temperatures.
A cutout from the central part of the observed field of view with the same extent as the here presented synthetic observations (i.e., 24\,Mm\,$\times$\,24\,Mm) is shown in Fig.~\ref{fig:Molnar}b to allow for a direct comparison.
Within the  1 and 99 percentile, the brightness temperature range is observed to be, between 5~kK and 11~kK, in the observational data is very similar to the range found for the simulated data (see Fig.~\ref{fig:H_a_ALMA_3mm_comparison_leenaarts_formula}e), 
in particular given the aforementioned uncertainties in absolute brightness temperatures derived from ALMA observations. This qualitative agreement with observations was already demonstrated for a different snapshot of the same Bifrost simulation by  \citet{2015A&A...575A..15L}. See also 
\citet{2004A&A...419..747L}. As mentioned in Sect.~\ref{sec:res_spatres}, the average brightness temperature at 3.0\,mm for the whole computational box of the 3D model is only $\sim 100$\,K  higher than the corresponding value for the VAL~C model and only 520\,K lower than the FAL~C value, which thus all agree with the reference value of 7300\,K derived by \citet{2017SoPh..292...88W} from ALMA observations within the aforementioned uncertainties.

\begin{figure*}[ht!]
\centering
\vspace*{-15mm}
\includegraphics[width=1.0\textwidth]{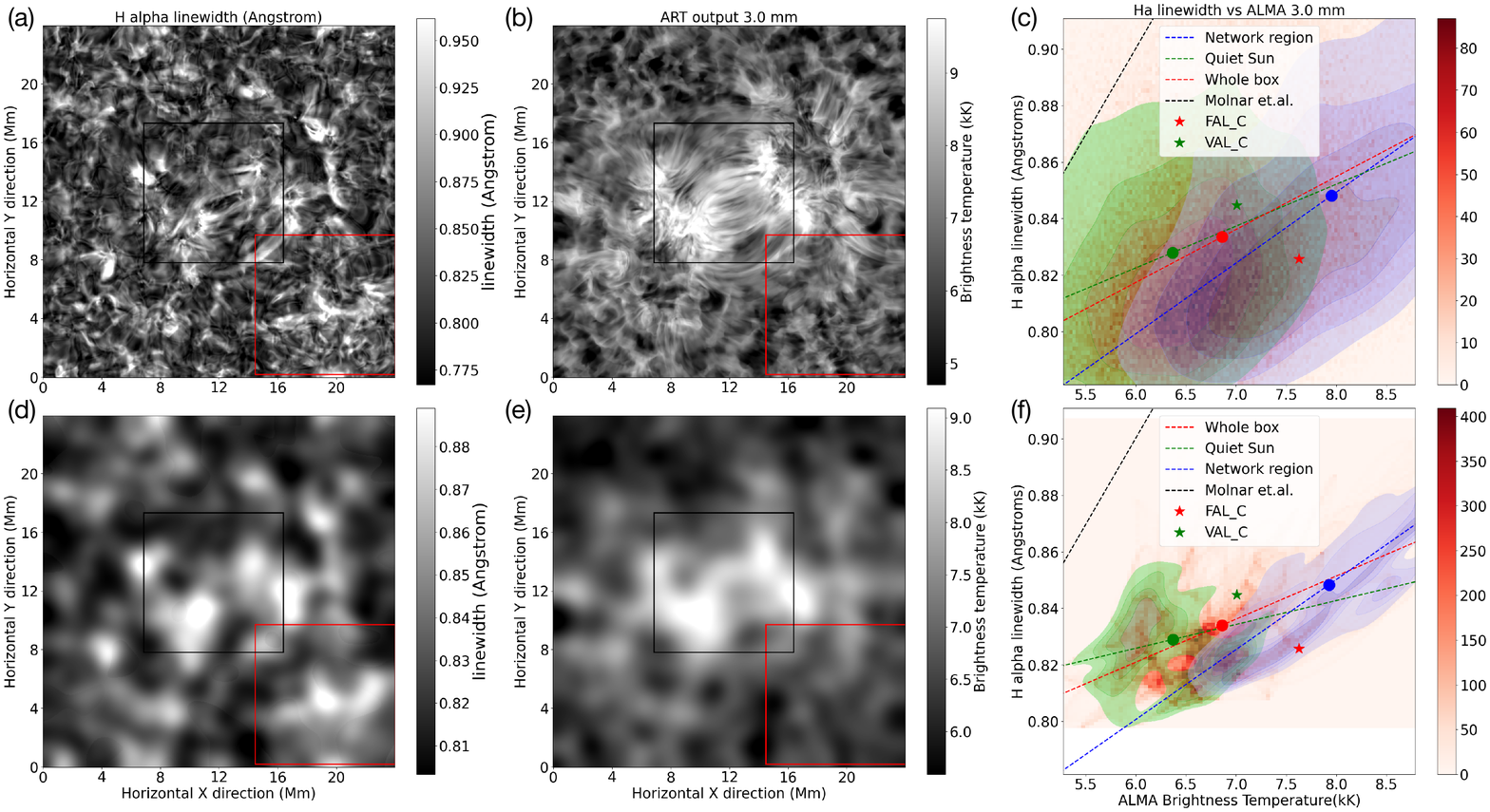}
\vspace*{-20mm}
\caption{Same as Fig.~\ref{fig:H_a_ALMA_3mm_comparison_leenaarts_formula_time_averaged}, for the  \citet{2019ApJ...881...99M} definition of linewidth. The black dashed line denotes the linear fit by  \citet{2019ApJ...881...99M}. }
\label{fig:H_a_ALMA_3.0mm_comparison_time_averaged}
\end{figure*}

\begin{figure*}[ht!]
\centering
\vspace*{-15mm}
\includegraphics[width=1.0\textwidth]{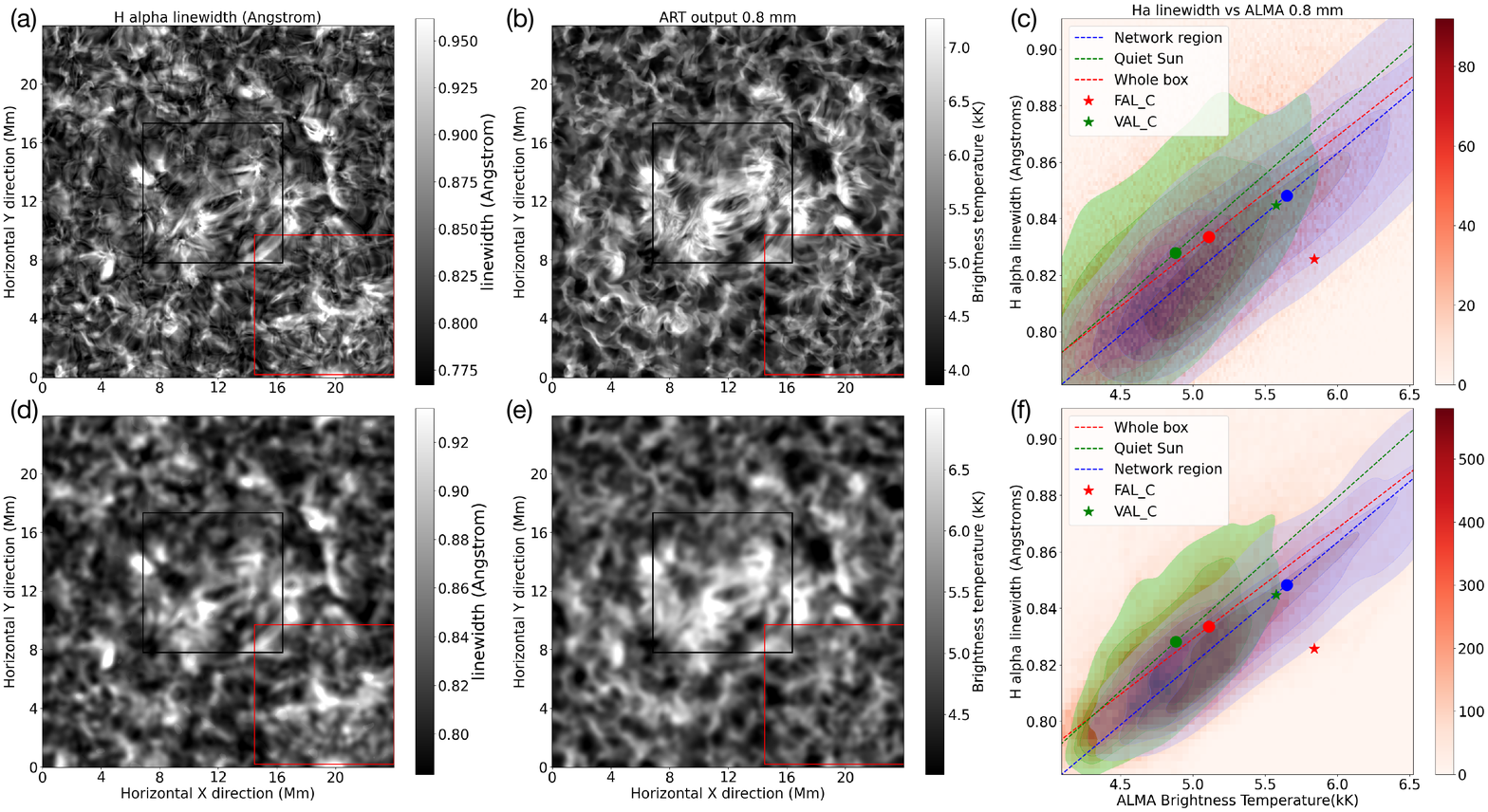}
\vspace*{-20mm}
\caption{Same as Fig.~\ref{fig:H_a_ALMA_3.0mm_comparison_time_averaged}, for 0.8~mm.}
\label{fig:H_a_ALMA_0.8mm_comparison_time_averaged}
\end{figure*}

\begin{figure}
\vspace*{-5mm}
\centering
\includegraphics[width=0.5\textwidth]{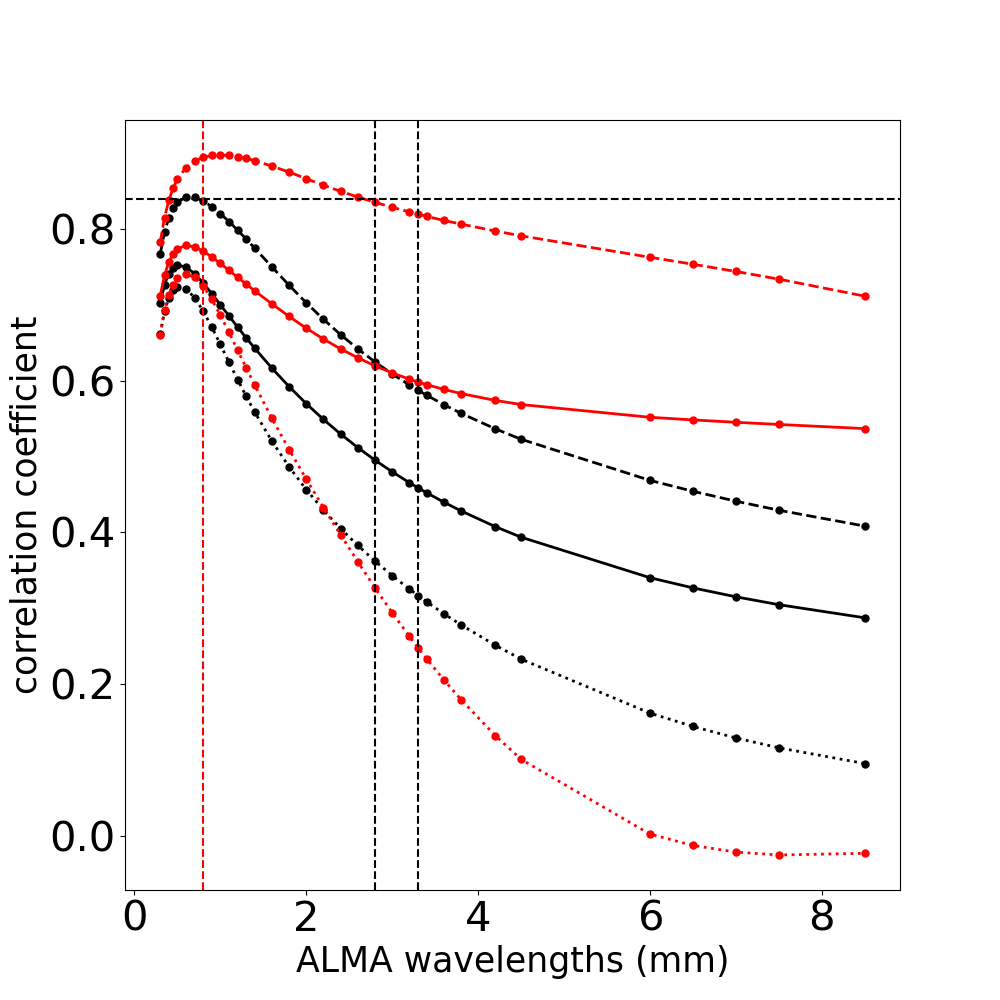}
\caption{Same as \ref{fig:H_a_ALMA_mm_correlations_Leenaarts_et_al}, for the time-averaged data set, for the  \citet{2019ApJ...881...99M} definition of linewidth. Legend same as Fig.~\ref{fig:H_a_ALMA_mm_correlations_Leenaarts_et_al}
\label{fig:H_a_ALMA_mm_correlations_time_averaged}}
\end{figure}

\begin{figure}
\vspace*{-2mm}
\centering
\includegraphics[width=0.5\textwidth]{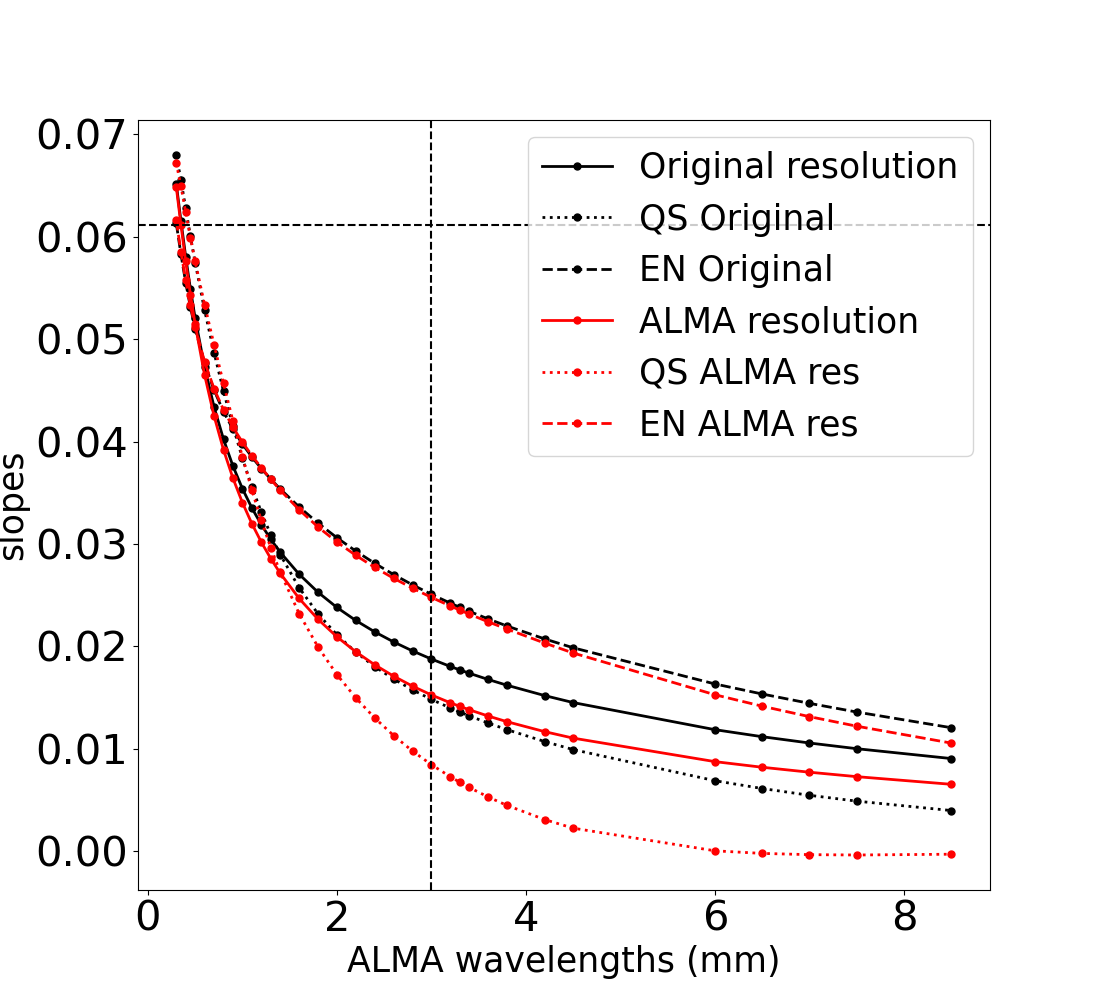}
\caption{Same as \ref{fig:H_a_ALMA_mm_slopes_Leenaarts_et_al}, for the time-averaged data set, for the  \citet{2019ApJ...881...99M} definition of linewidth.}
\label{fig:H_a_ALMA_mm_slopes_time_averaged}
\end{figure}

\begin{figure*}[t!]
    
    \vspace*{-13mm}
    \includegraphics[width=1.0\textwidth]{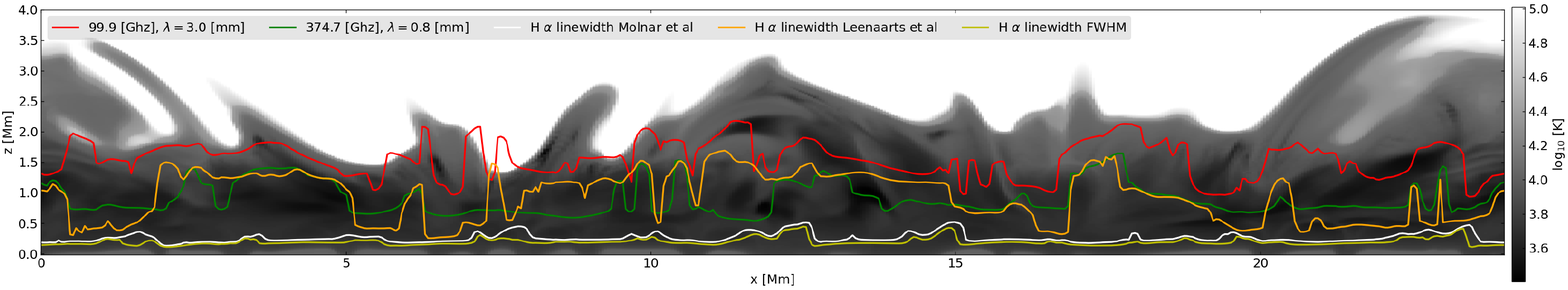}
    \vspace*{-87mm}
   \caption{ 
  Simulated logarithmic gas temperature (grey-scale) in a vertical slice of 3D rMHD Bifrost simulations. The solid lines represent the heights at which the optical depth is unity at a wavelength of 3.2~mm (red), and 0.8~mm (green), which correspond to ALMA bands 3, 5 and 7, respectively. The orange, white and yellow lines show the average heights at which the red and blue sides of the H$\alpha$ line width according to the \citet{2012ApJ...749..136L}, \citet{2019ApJ...881...99M} and FWHM definitions of linewidth are formed.}
\label{fig:formation_heights}
\end{figure*}

\begin{figure*}[ht!]
\vspace*{-15mm}
\includegraphics[width=1.0\textwidth]{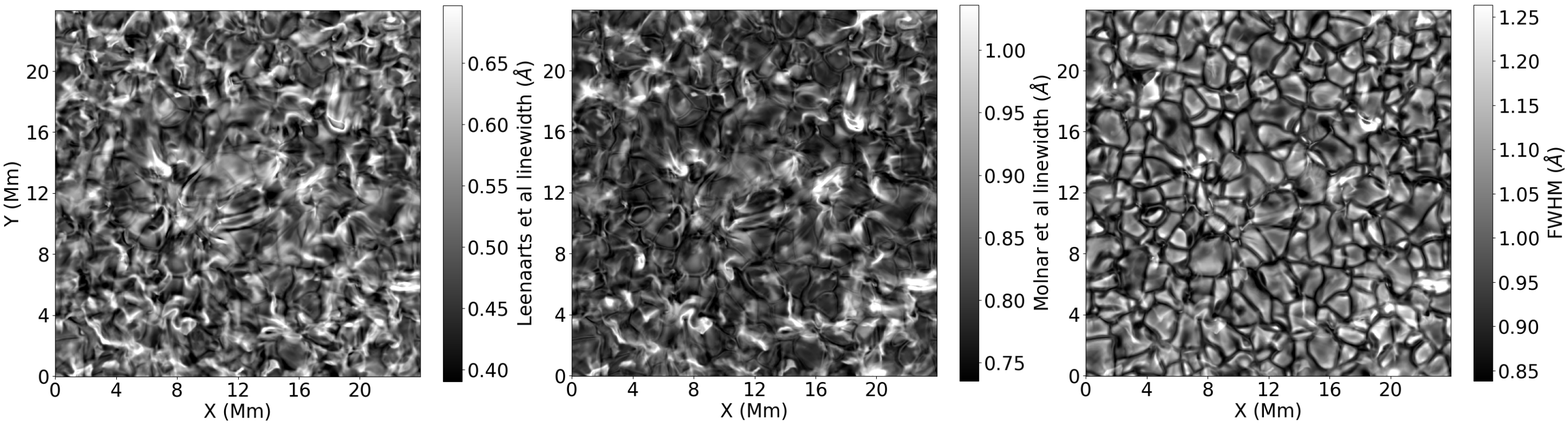}
\vspace*{-73mm}
\caption{Maps of linewidths for original resolution case for the single snapshot: Left: \citet{2012ApJ...749..136L} definition, Middle: \citet{2019ApJ...881...99M} definition and Right: FWHM.}
\label{fig:Linewidths_three_maps}
\end{figure*}

\begin{figure*}[t!]
\vspace*{-10mm}
\includegraphics[width=0.52\textwidth]{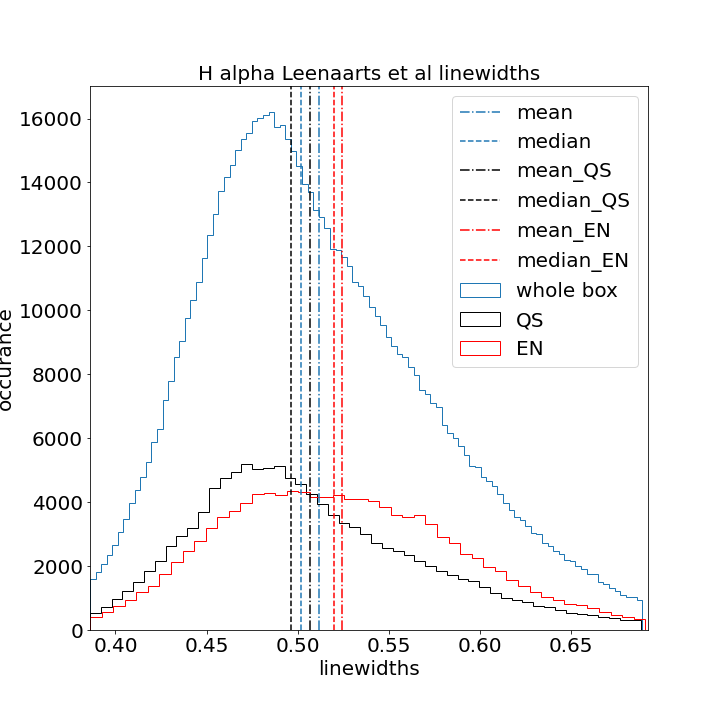}
\includegraphics[width=0.52\textwidth]{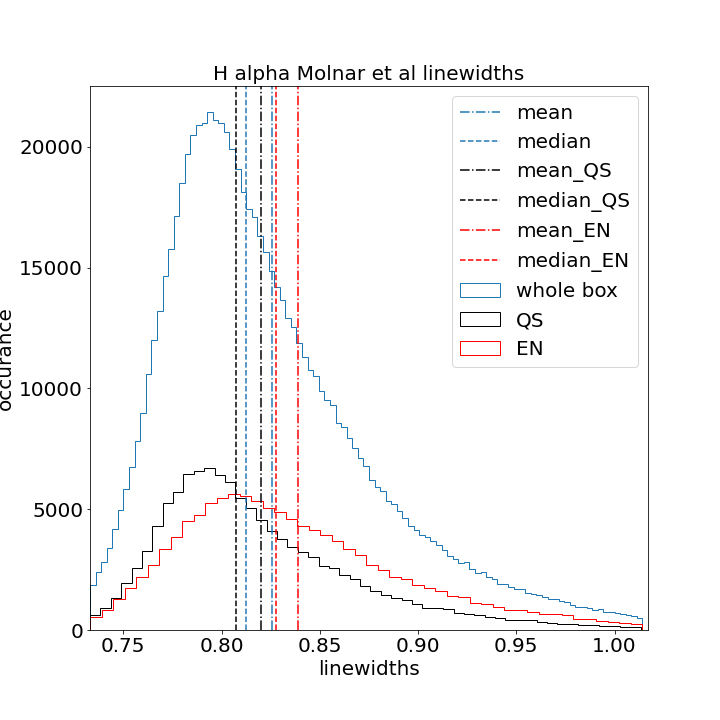}
\vspace*{-10mm}
\caption{Left: Distribution of  H$\alpha$ line core widths \citep[following the definition by][] {2012ApJ...749..136L} for the whole time series. Right: 
Corresponding line widths according to the definition by \citet{2019ApJ...881...99M}. The whole box, QS and EN regions are separately plotted in blue, black and red, and mean and median values for the three sets of pixels are indicated with a dot-dashed and dashed lines in the respective colours. Please note the difference in ranges for linewidths axes in the two plots.}
\label{fig:Linewidths_leenaarts_histogram}
\end{figure*}

The resolution of the observed ALMA data shown in Fig.~\ref{fig:Molnar} appears to be lower than for the degraded maps presented in Fig.~\ref{fig:H_a_ALMA_0.8mm_comparison_leenaarts_formula}e, which can be attributed to several factors.
Firstly, as detailed in Sect.~\ref{sec:psf}, the synthetic brightness temperatures maps are degraded to ALMA's angular resolution by convolution with ideal PSFs, each of which is constructed as a circular Gaussian with a width appropriate for the respective wavelength. 
This simplified approach corresponds in principle to a telescope with a filled aperture of the size of the longest baseline.  
In reality, however, ALMA consists of a limited number of antennas in a given configuration, which results in a sparse sampling of the spatial Fourier space of the observed source.   
The resulting PSF (or rather referred to as synthesised beam) of the interferometric array is typically elongated and tilted.  
Properly accounting for these detailed instrumental effects, the additional degradation due to Earth's atmosphere and the necessary imaging of the ALMA data \citep[see, e.g.,][]{2022A&A...659A..31H} requires a much more detailed and computationally more expensive modelling approach \citep{wedemeyerfrontiers}, which goes beyond the scope of the study presented here. Overall, however, it is safe to assume that the employed degradation of the synthetic brightness temperature maps is still too optimistic. The same is true for the synthetic H$\alpha$ data for which further instrumental effects and the influence of the Earth's atmosphere are not accounted for. 
The statement that the employed instrumental degradation for ALMA is still too optimistic is confirmed by the power spectral density shown for the ALMA Band~3  in Fig.~\ref{fig:PSD_t_avg}. The curve for the observational 
data starts to drop at a slightly larger spatial scale as compared to the simulated data and also shows a noise component at smaller spatial scales that is not included in the simulations here.

Secondly, the ALMA image shown in Fig.~\ref{fig:Molnar} represents the time-average over an observational period of 10~min that is produced in the same way as described by \citet{2019ApJ...881...99M}. 
Also, the H$\alpha$ observations obtained with IBIS by \citeauthor{2019ApJ...881...99M} were time-averaged for a comparison with the aforementioned ALMA data.  
A period of 10\,min is substantial given the dynamic time scales on which the chromospheric fine structure evolves, the time-averaging further blurs the resulting images. 
This effect can be seen by comparing, e.g., Fig.~\ref{fig:H_a_ALMA_3mm_comparison_leenaarts_formula}e and Fig.~\ref{fig:H_a_ALMA_3.0mm_comparison_time_averaged}e but it is also apparent in Fig.~3 by \citeauthor{2019ApJ...881...99M}, which compares snapshots to time-averaged images. 
In Table~\ref{table:all_three_combined}, the correlation coefficients and the slopes for both cases are listed. 

Thirdly, limited spectral sampling points for the IBIS observations \citep{2006SoPh..236..415C} and also the time it takes to scan through the spectral line profile would contribute to uncertainties. For our simulated data, we have used more than three times more spectral sampling points for the H$\alpha$ line, namely 101, as compared to the 29 points for the IBIS data used by \citet{2019ApJ...881...99M}. 

As a result of the limited resolution discussed above, emission from different heights gets mixed within the respective synthesised ALMA beam. This effect is particularly important to be aware of in view of the highly dynamic and intermittent nature of the solar chromosphere, which can result in substantial changes in formation height across small (possibly unresolved) spatial distances.  
In addition, it should be noted that
the region observed by  \citet{2019ApJ...881...99M} comprises a mix of plage, network and magnetically quiet regions, whereas the simulated data contains only EN and QS regions and can thus only be compared within the respective limitations.  

As a result of these effects, differences in the density plot for H$\alpha$ linewidth vs. ALMA brightness temperature for the simulated data in Fig.~\ref{fig:H_a_ALMA_3.0mm_comparison_time_averaged}f and the corresponding density plot for observed data in \citet[][see their Fig.~4]{2019ApJ...881...99M} are expected. It should be noted though that, despite these effects, the distributions in observed and simulated data are very similar. The simulated QS distribution coincides with the high-density scatter distribution at low observed brightness temperatures and low H$\alpha$ line widths in the plot shown by \citeauthor{2019ApJ...881...99M}.  The blue EN contour, which has a slightly lower slope than the slopes for the QS part and the slope for the whole box in the simulated data, has its counterpart in the scatter plot for the observed data. The calculated slopes for the whole box at the degraded resolution are slightly different, which can be attributed to the aforementioned differences between the observations and the simulations reported here. For the observed data, \citet{2019ApJ...881...99M} report the slope to be 0.0612 as compared to 0.015 for the whole box at degraded spatial and temporal resolution for the simulated data (see Table~\ref{table:all_three_combined} for details).

As shown in Sect.~\ref{sec:correlations_and_slopes}, changing the spatial resolution affects the correlations between H$\alpha$ linewidth and mm brightness temperatures. The correlation for the data observed at 3\,mm reported by \citet{2019ApJ...881...99M} is 0.84, which is close to the correlation of 0.61 found here for the simulated data at degraded spatial and temporal resolution. The correlation is 0.5 for the data at the original resolution. The difference in the correlation coefficients is because the spatial arrangement of the individual pixels relative to each other does not matter when calculating them. Consequently, the correlation coefficient increases with decreasing spatial resolution as the resulting image degradation gradually removes the largest variations on the smallest scales in both the ALMA and the H$\alpha$ maps. For both data sets, lowering the spatial resolution results in narrower intensity distributions towards the mean value.

The height at which the core is formed is higher and the linewidth lower for the quiet region and the formation height is lower and linewidth is higher for the EN region \citep{2012ApJ...749..136L}. 

\subsection{Effect of temporal resolution}
\label{sec:discuss_temp_res}

When time-averaging the spatially degraded H$\alpha$ and mm~brightness temperature maps, the resulting correlation coefficients change with respect to the results based on single snapshots. 
As summarised in Table~\ref{table:all_three_combined}, for the line core width as defined by \citet{2012ApJ...749..136L}, the correlation coefficients for all three pixel sets become lower when time-averaging both at original and reduced resolution except for a small increase for the correlation with the 3\,mm data across the whole box at reduced resolution  (cf.  Fig.~\ref{fig:H_a_ALMA_mm_correlations_time_averaged}). 
The same behaviour is found when using the linewidth definition by \citet{2019ApJ...881...99M}, maybe except for a small increase of the EN correlations both at original and reduced spatial resolution. 

In general, temporal averaging results in blurring of the maps and  lower correlation coefficients. 
For the EN region, however, time-averaging can tend to highlight persistent features, resulting in a slightly higher correlation coefficient. 
In contrast, for the QS region, the more uniformly distributed and short-lived features get averaged over time and space and lose the similarities. Similarly, the values of slopes for original and degraded resolutions are very close for individual wavelengths (see  Fig.~\ref{fig:H_a_ALMA_mm_slopes_time_averaged}). 
In conclusion, time averaging causes a decrease in the correlation in the QS case and a small increase in the EN case for the Molnar definition while the slopes decrease marginally. 
This effect becomes slightly more notable for the maps at reduced spatial resolution as compared to the maps at original resolution. We conclude that the exact values and behaviour of correlation coefficients as shown here may depend critically on the spatial and temporal resolution of the employed data.}

\subsection{Comparison of linewidth definitions}

As seen in Fig.~\ref{fig:line_profiles_linewidth_definition_comparison}, the value and the atmospheric layer connected to the linewidth change significantly based on the definition used. In the case of the \citet{2019ApJ...881...99M} definition, as described in Sect.~\ref{sec:deflinewidth}, the continuum and thus the line depth are derived from the $\pm 1$\,\AA~ region around the line core. 
On the other hand for both the \citet{2012ApJ...749..136L} definition and the FWHM, the continuum is defined based on the true continuum intensity. The wavelength grid points used for the calculation of the line core width, which are close to the line core, are formed significantly higher in the atmosphere as compared to the \citeauthor{2019ApJ...881...99M} approach as illustrated in
Fig.~\ref{fig:formation_heights}. The radiation at the wavelength points in line wings as used for the FWHM definition are accordingly formed lower in the atmosphere, even lower than for the \citeauthor{2019ApJ...881...99M} definition. 
In Fig.~\ref{fig:Linewidths_three_maps}, the line widths for all three definitions are compared, implying the above-discussed differences in connected atmospheric heights, being highest for the line core width definition by  \citet{2012ApJ...749..136L}. The FWHM linewidth map much resembles a granulation pattern as formed in the low photosphere.

The distribution of line core widths as seen in Fig.~\ref{fig:Linewidths_leenaarts_histogram}a is in the range of line core widths calculated by \citet{2012ApJ...749..136L}. Using the  \citet{2019ApJ...881...99M} definition instead  produces larger linewidths (see Fig.~\ref{fig:Linewidths_leenaarts_histogram}b) which nonetheless remain smaller than the observationally determined values reported by \citet{2019ApJ...881...99M} . 
The simulated linewidths according to the \citet{2019ApJ...881...99M} definition cover a range of 0.75 to 1.0.\,\AA, whereas the values derived from observations by \citet{2019ApJ...881...99M} are in the range of 0.9 to 1.3\,\AA, which is also in line with the observed linewidth for $\alpha$~Cen~A (see Sect.~\ref{sec:alf_sec_A}). 
The reason behind this discrepancy could be that the H$\alpha$ opacity used in the simulations including the radiative transfer calculations is lower than the real Sun \citep{2012ApJ...749..136L}. Similarly, as seen in Fig.~\ref{fig:formation_heights}, the formation heights of the radiation at the wavelength positions entering the line core width calculation with the \citet{2012ApJ...749..136L} definition are closer to the formation heights of the mm continua as compared to the corresponding heights for the  \citet{2019ApJ...881...99M} definition.
The conclusion that the \citet{2012ApJ...749..136L} definition produces a better match of the H$\alpha$ line (core) width with the mm continua is also implied by a closer match in terms of spatial power spectral density as compared to the \citet{2019ApJ...881...99M} definition (see Fig.~\ref{fig:PSD_t_avg}). 
This seems to be true at the original and at the reduced resolution. 

\subsection{Implications for solar-like stars}
\label{sec:alf_sec_A}
The correlation between H$\alpha$ linewidth and mm brightness temperatures described in Sect.~\ref{sec:correlations_and_slopes} can be transferred to the study of sun-like stars.
It should be noted that the study presented here is based on simulated EN and QS regions, while the corresponding trends for a whole active region could not be investigated due to the lack of an equivalent consistent simulation. Any comparison with unresolved observations of other solar-like stars, therefore, needs to account for the potential contributions of active regions via corresponding filling factors.

A direct application to stellar observations is to estimate the stellar activity level, in particular for sun-like stars as for instance used in the studies by \citet{2016A&A...594A.109L} and \citet{2021A&A...655A.113M, 2022A&A...664L...9M}.
The position in the H$\alpha$ linewidth - mm brightness temperature plane as derived from adequate observations then provides constraints for the properties of the observed stellar atmosphere. 
This approach is illustrated by comparing the data points for the semi-empirical 1D models FAL~C and VAL~C to the averages for the whole box, EN and QS data points on the H$\alpha$ linewidth - mm brightness temperature plane in Figs.~\ref{fig:H_a_ALMA_3mm_comparison_leenaarts_formula}~c,f and \ref{fig:H_a_ALMA_0.8mm_comparison_leenaarts_formula}~c,f. 
For the disk-integrated signal, the centre-to-limb variation for the H$\alpha$ line and ALMA full disk data should be taken into account \citep[see, e.g.,][and references therein]{2022FrASS...9.1320A, 2022A&A...661L...4A, 2022ApJ...939...98O,2018A&A...619L...6N}. As the H$\alpha$ core forms in the chromosphere and the wings in the photosphere, the limb effects would be varying (brightening or darkening depending on the wavelength) when going from the core to the wing. In a recent study based on SST observations by \citet{2022arXiv221203991P}, they concluded that chromospheric lines might exhibit a blue shift towards the limb due to the chromospheric canopies, but the resulting effect on the equivalent width or linewidths is nontrivial. Further, they show that the line core widths would decrease when going from the centre to limb. More systematic studies need to be conducted in order to understand the limb effects on the correlations between the two diagnostics and the implications for disk-integrated stellar observations.

The brightness temperature values at the wavelengths of 0.8~mm and 3~mm for $\alpha$~Cen~A, which are obtained from ALMA observations \citep[see][and references therein]{2021A&A...655A.113M}, are in line with the values derived from ALMA observations of the Sun and also with those based on the simulations presented in this study. In contrast, the H$\alpha$ line profile for $\alpha$~Cen~A (see Fig.~\ref{fig:line_profiles_linewidth_definition_comparison}) is with a line core width of $\sim 0.6$\,\AA~ significantly broader than the simulated data based on the Bifrost simulation and also for the FAL~C and VAL~C models, which causes the respective data point to be placed off the linear fit for the solar simulation shown in Figs.~\ref{fig:H_a_ALMA_3mm_comparison_leenaarts_formula}c,f and \ref{fig:H_a_ALMA_0.8mm_comparison_leenaarts_formula}c,f 
(see the black circle with error bars). 
The  $\alpha$~Cen~A datapoint is out of the bounds of the plane in Figs.~\ref{fig:H_a_ALMA_3.0mm_comparison_time_averaged}~c,f and \ref{fig:H_a_ALMA_0.8mm_comparison_time_averaged}~c,f as the calculated \citet{2019ApJ...881...99M} linewidth is $~1.1 \AA$. 
The comparatively broad H$\alpha$ line profile for $\alpha$~Cen~A is likely due to differences in atmospheric structure and the activity level of this star as compared to the Sun although an in-depth study would be needed to answer this in detail. We also note that the solar observations are spatially resolved and cover a small region on the Sun only, whereas the $\alpha$~Cen~A data intrinsically integrated across the whole (spatially unresolved) stellar disk.

\section{Conclusions}
\label{Sec:Conclusion}

We look at the synthetic H$\alpha$ linewidth and the brightness temperatures corresponding to the mm wavelengths generated from a 3D close to realistic Bifrost model. Degrading the spatial resolution of these synthetic observations to ALMA resolution results in decreased standard deviations of the synthetic observables: H$\alpha$ line width and mm brightness temperatures, and increased correlations between them. With increasing wavelength, the beam size increases and the small structures get blurred out, resulting in a further increased correlation between the two diagnostics. But this degradation in terms of spatial resolution does not have much effect on the mean slopes for the linear fits on the H$\alpha$ linewidth vs mm brightness temperature distributions. This implies that the instrumental resolution of ALMA is sufficient to capture the similarities between the two diagnostics.

The solar results in comparison with the observations of $\alpha$~Cen~A strongly imply that the H$\alpha$ linewidth and mm continuum brightness temperatures can be considered equivalent indicators, as seen in Fig.~\ref{fig:formation_heights} that they form close to each other and that the thermal stratification of a star can be constrained using the correlations and slopes found in this study in the absence of mm data, by just using the H$\alpha$ data and vice versa. As discussed in \citet[][see also references therein]{2016SSRv..200....1W}, using multi-wavelength observations along with mm observations can provide better insights into the thermal structure of stellar atmospheres.

The H$\alpha$ linewidth depends on the temperature in the line-forming region which is observed using mm continua \citep{2012ApJ...749..136L}. The brightness temperatures can be directly calculated from the mm continuum intensity maps which correlate with the H$\alpha$ linewidth very well. The best match in terms of highest correlation is found for a wavelength of 0.8~mm, which corresponds to ALMA Band~7. 
Simultaneous H$\alpha$--ALMA Band~7 observations have therefore the potential advantage of better constraining the imaging process and thus resulting in more reliable temperature measurements for the solar chromosphere. 
Published simulations of the solar chromosphere \citep{2016A&A...585A...4C} show too narrow MgII h \& k lines, compared with IRIS observations \citep{2019ARA&A..57..189C}, indicating that the models have too little mass at upper chromospheric temperatures. Newer models in higher resolution including ambipolar diffusion and non-equilibrium ionisation \citep[e.g.,][]{2023ApJ...943L..14M}, flux emergence \citep{2023ApJ...944..131H} or run with the MURaM code with chromospheric extensions \citep{2022A&A...664A..91P} show significantly broader magnesium lines. It will be important to check the reported ALMA-H$\alpha$ relationships with these new models.
As the correlation between the two diagnostics is high even in the case of lower resolution, this study may prove useful in mitigating the challenging task of calibration of mm continuum solar data obtained from ALMA.

\begin{acknowledgements}
We thank the referee for very valuable comments and suggestions, which improved considerably the quality of this paper. This work is supported by the Research Council of Norway through the EMISSA project (project number 286853), the Centres of Excellence scheme, project number 262622 (``Rosseland Centre for Solar Physics'') and through grants of computing time from the Programme for Supercomputing.  
This paper makes use of the following ALMA data: ADS/NRAO.ALMA\#2016.1.01129.S. ALMA is a partnership of ESO (representing its member states), NSF (USA) and NINS (Japan), together with NRC (Canada), MOST and ASIAA (Taiwan), and KASI (Republic of Korea), in cooperation with the Republic of Chile. The Joint ALMA Observatory is operated by ESO, AUI/NRAO and NAOJ. The authors thank \citeauthor{2008A&A...488..653P} for their prompt correspondence regarding the H$\alpha$ observational data of $\alpha$ Cen A.
This research utilised the Python libraries matplotlib \citep{2007CSE.....9...90H}, seaborn \citep{Waskom2021} and the NumPy computational environment \citep{citeulike:9919912}. The authors would like to thank Thore Espedal Moe, Mats Ola Sand, Atul Mohan, Juan Camilo Guevara Gomez, Vasco Henriquez, Shahin Jafarzadeh for their support.

\end{acknowledgements}

\bibliographystyle{aa}
\bibliography{main}

\begin{appendix}

\onecolumn
\section{Additional brightness temperature maps for single snapshot with  \citet{2012ApJ...749..136L} linewidth}
\label{Appendix_B}

\begin{figure*}[hbt!]

\centering
\includegraphics[width=1.0\textwidth]{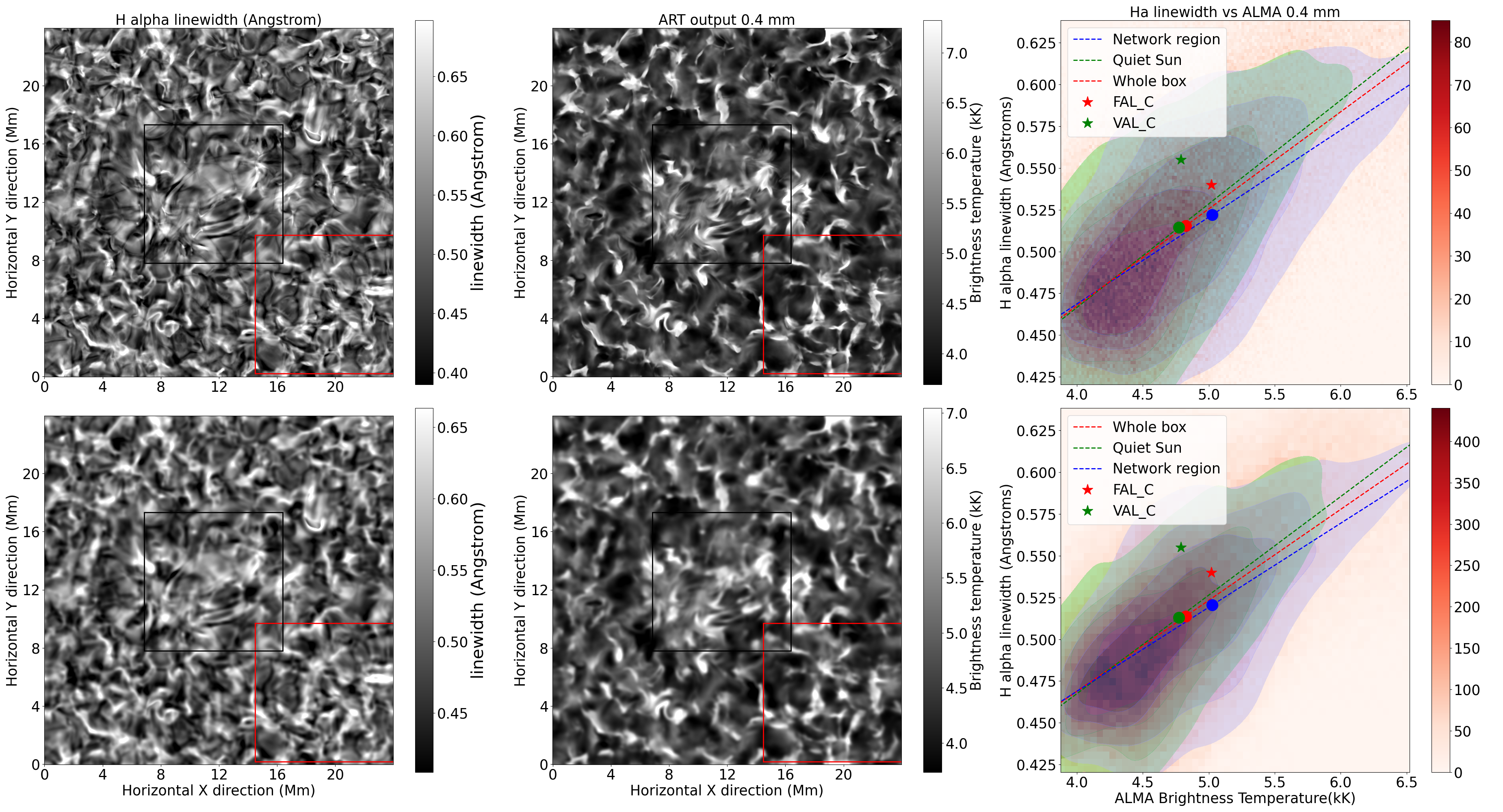}
\caption{Same as Fig.~\ref{fig:H_a_ALMA_3mm_comparison_leenaarts_formula}, for 0.4\,mm corresponding to ALMA Band 9-10.}
\label{fig:H_a_ALMA_0.4mm_comparison} 

\end{figure*}

\begin{figure*}[hbt!]
\centering
\includegraphics[width=1.0\textwidth]{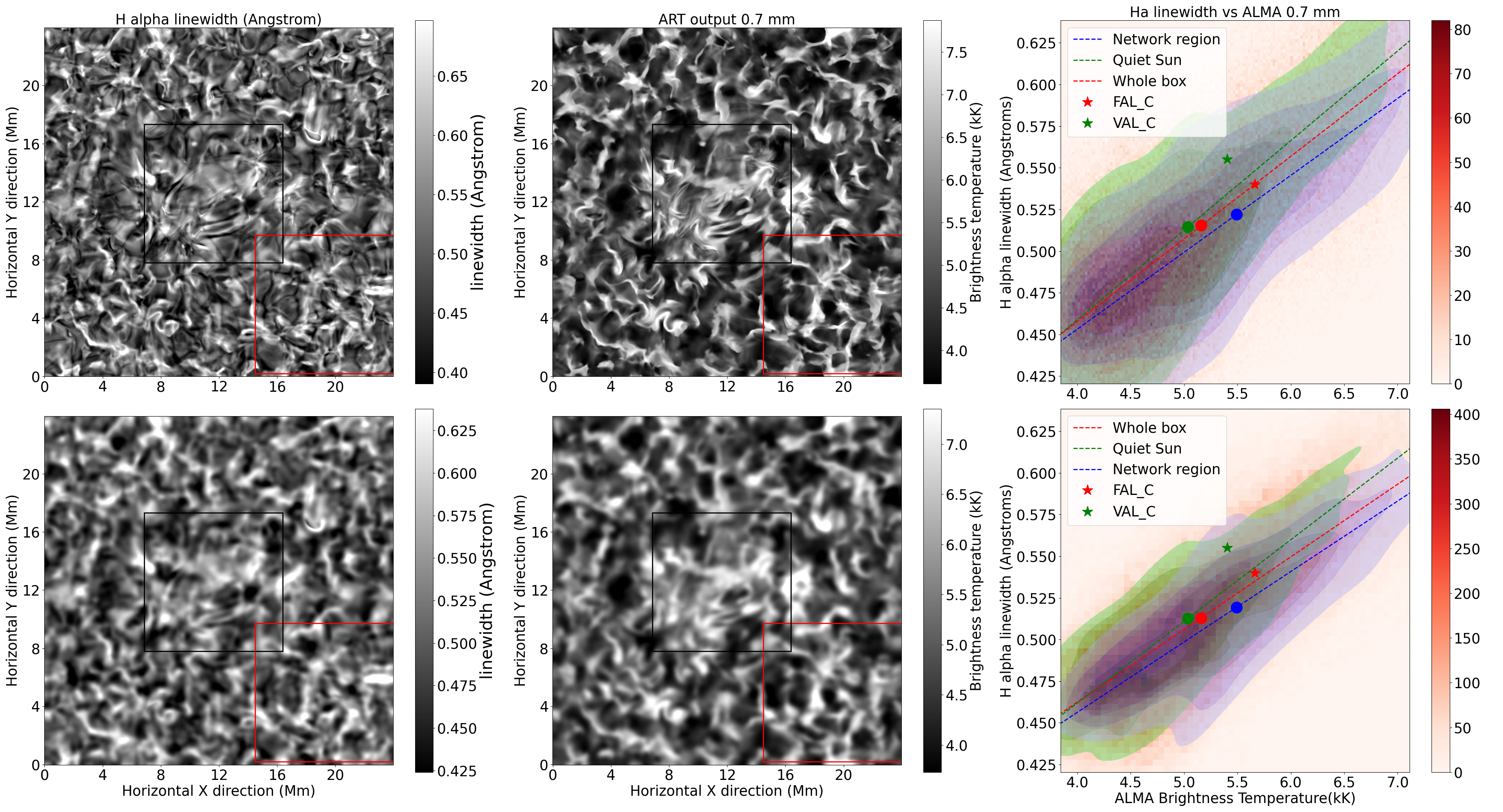}
\caption{Same as Fig.~\ref{fig:H_a_ALMA_3mm_comparison_leenaarts_formula}, for 0.7\,mm corresponding to ALMA Band 8.}
\label{fig:H_a_ALMA_0.7mm_comparison}
\end{figure*}

\begin{figure*}
\centering
\includegraphics[width=1.0\textwidth]{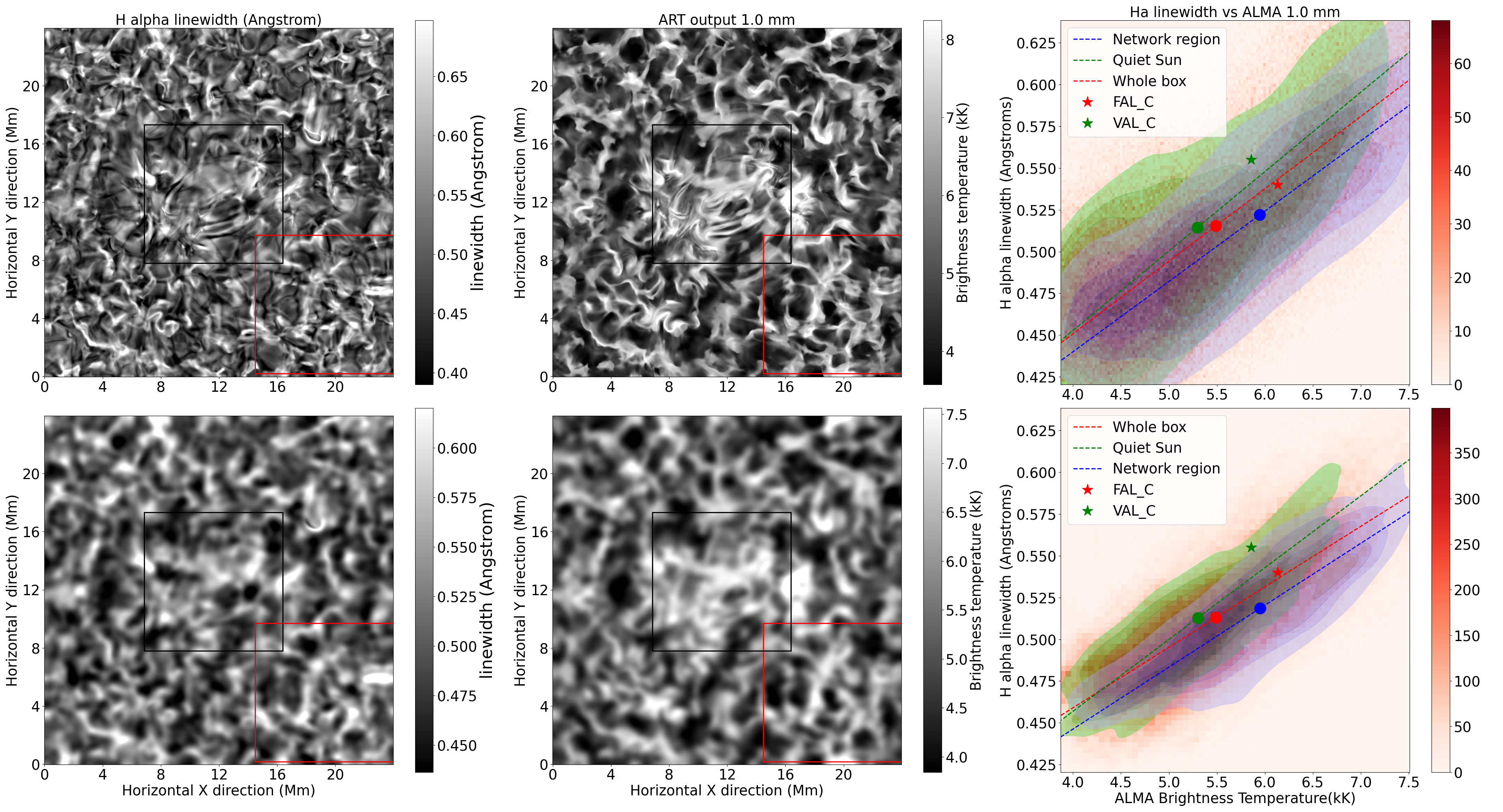}
\caption{Same as Fig.~\ref{fig:H_a_ALMA_3mm_comparison_leenaarts_formula}, for 1.0\,mm corresponding to ALMA Band 7.}
\label{fig:H_a_ALMA_1.0mm_comparison}
\end{figure*}

\begin{figure*}
\centering
\includegraphics[width=1.0\textwidth]{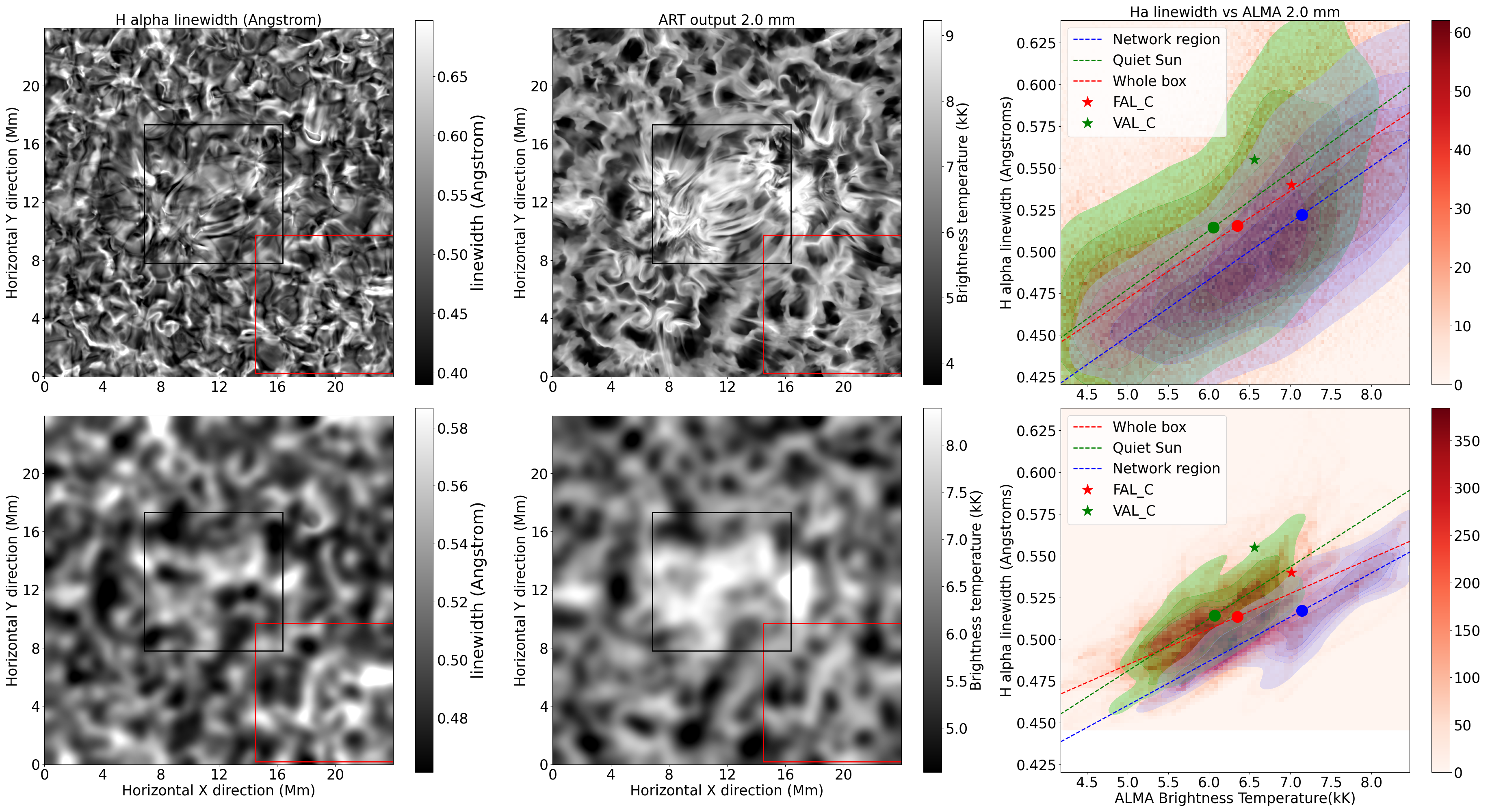}
\caption{Same as Fig.~\ref{fig:H_a_ALMA_3mm_comparison_leenaarts_formula}, for 2.0\,mm corresponding to ALMA Band 4}.
\label{fig:H_a_ALMA_2.0mm_comparison}
\end{figure*}

\begin{figure*}
\centering
\includegraphics[width=1.0\textwidth]{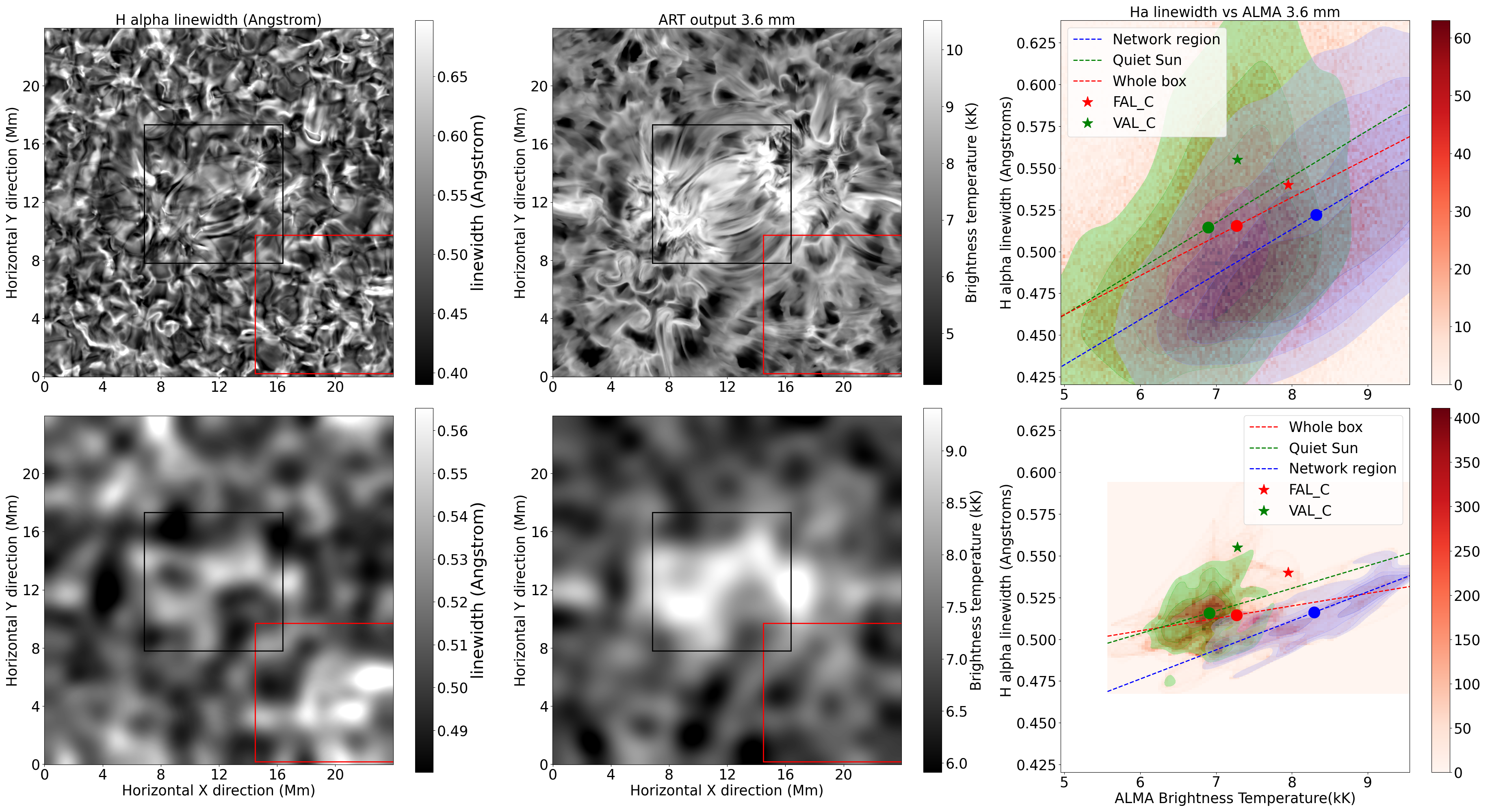}
\caption{Same as Fig.~\ref{fig:H_a_ALMA_3mm_comparison_leenaarts_formula}, for 3.6\,mm corresponding to ALMA Band 3.}
\label{fig:H_a_ALMA_3.6mm_comparison}
\end{figure*}

\begin{figure*}
\centering
\includegraphics[width=1.0\textwidth]{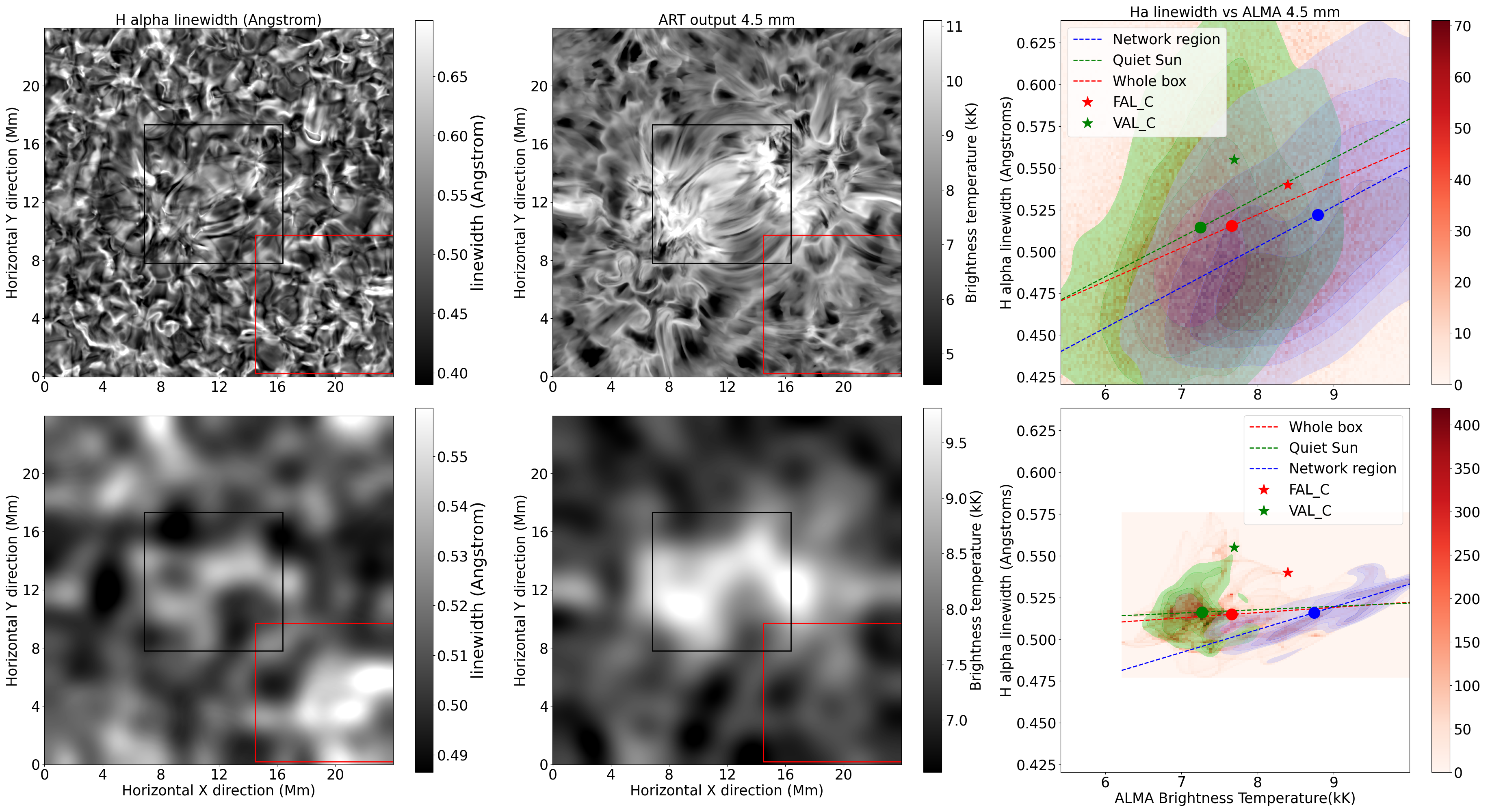}
\caption{Same as Fig.~\ref{fig:H_a_ALMA_3mm_comparison_leenaarts_formula}, for 4.5\,mm corresponding to ALMA Band 2.}
\label{fig:H_a_ALMA_4.5mm_comparison}
\end{figure*}

\begin{figure*}
\centering
\includegraphics[width=1.0\textwidth]{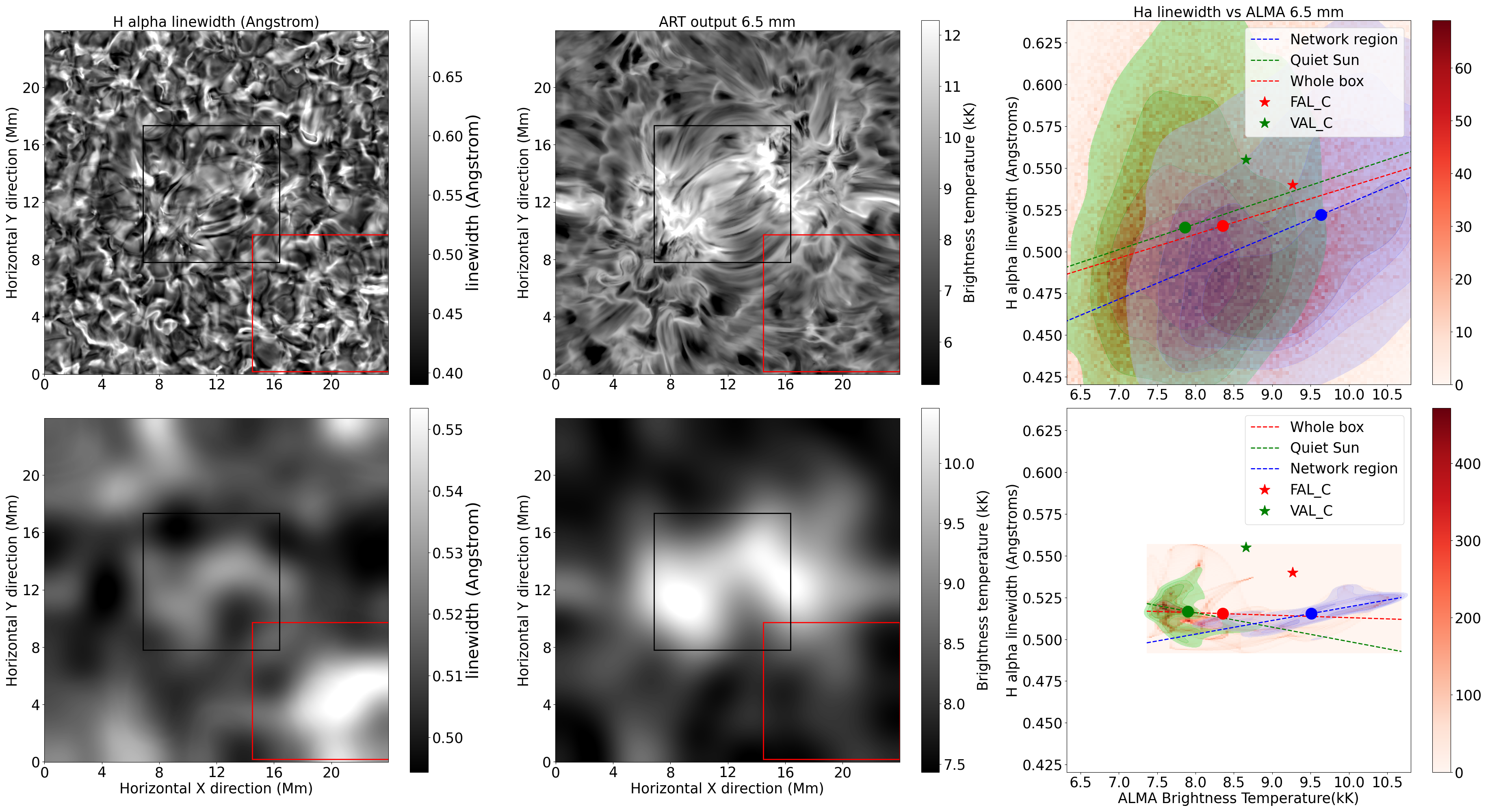}
\caption{Same as Fig.~\ref{fig:H_a_ALMA_3mm_comparison_leenaarts_formula}, for 6.5\,mm corresponding to ALMA Band 1.}
\label{fig:H_a_ALMA_6.5mm_comparison}
\end{figure*}

\begin{figure*}
\centering
\includegraphics[width=1.0\textwidth]{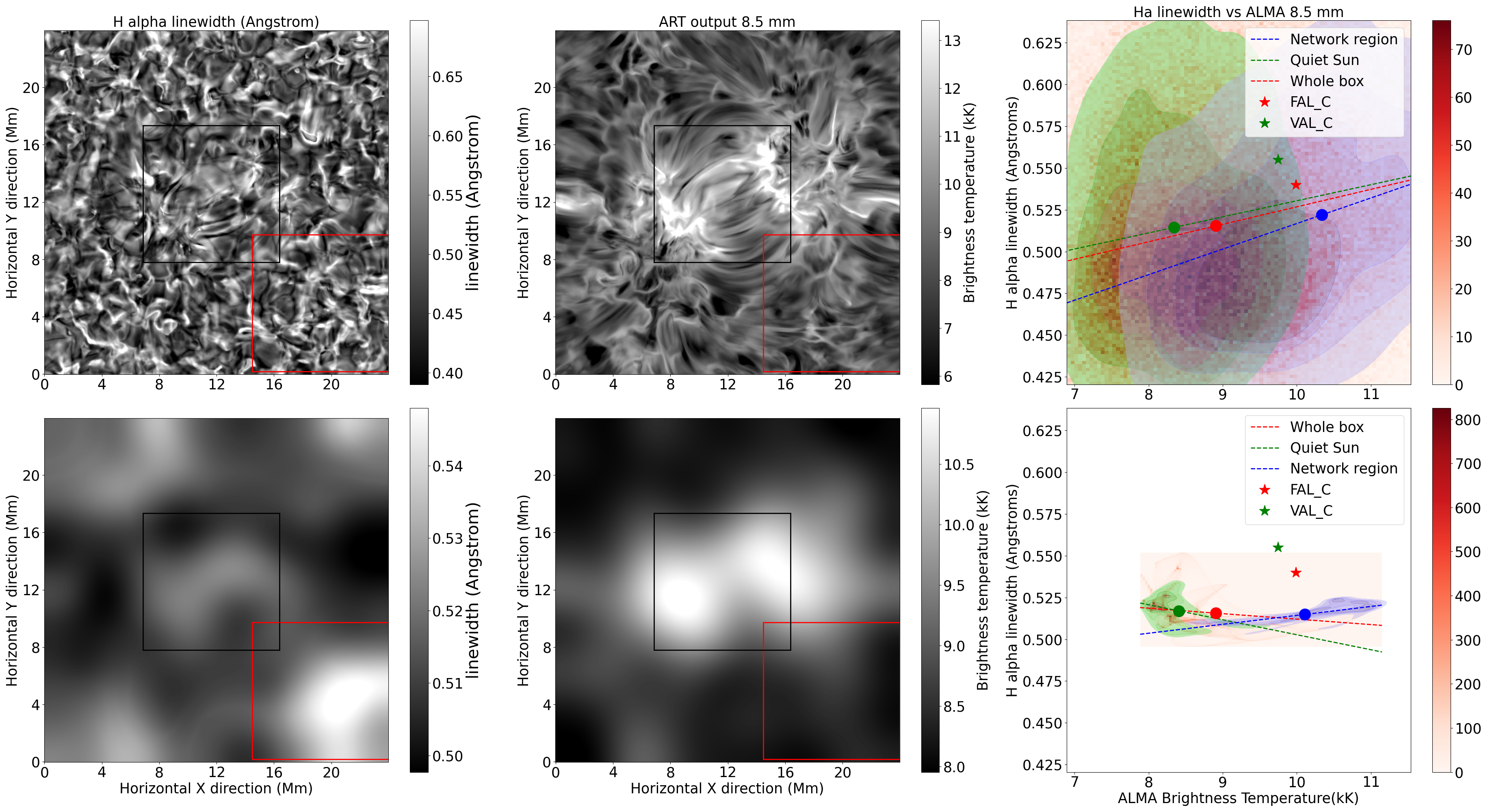}
\caption{Same as Fig.~\ref{fig:H_a_ALMA_3mm_comparison_leenaarts_formula}, for 8.5\,mm corresponding to ALMA Band 1.}
\label{fig:H_a_ALMA_8.5mm_comparison}
\end{figure*}

\end{appendix}

\end{document}